\documentstyle[psfig,aps,epsf]{revtex} 
\newcommand{\bc}{\begin{center}}
\newcommand{\ec}{\end{center}} 
\newcommand{\bi}{\begin{itemize}}
\newcommand{\ei}{\end{itemize}} 
\newcommand{\bt}{\begin{table}}
\newcommand{\et}{\end{table}} 
\newcommand{\beq}{\begin{equation}}
\newcommand{\eeq}{\end{equation}}

\newcommand{\SA}{ \mbox{({\bol A},{\bol S})}}
\newcommand{\bol}[1]{ \mbox{\boldmath $ #1 $} }
\newcommand{\C}{\cal C}

\begin{document}

\title{Protein folding using contact maps}
\author{Michele Vendruscolo and Eytan Domany}
\address{Department of Physics of Complex Systems, 
Weizmann Institute of Science, Rehovot 76100, Israel}

\maketitle

\tableofcontents

\section{Introduction}

Computational approaches to protein folding 
are divided into two main categories.
In {\it energy minimization} methods the native state is identified with the
the ground state of a suitable energy function (Brooks, 1998).
In {\it fold recognition} methods, the native state is selected
as the most compatible structure among those present in a library
(Bowie {\em et al.}, 1992; Jones {\em et al.}, 1992; Fisher {\em et al.}, 1996).
Both approaches depend on three important choices:
1) The representation of protein structure;
2) The set of alternative structures among which the native fold is sought for;
3) A bias towards the ``best'' conformation.
In energy minimization approaches such a bias is an energy function,
and the best conformation is the one of lowest energy.
On the other hand, in fold recognition methods, a compatibility
function for a sequence on a structure is used. The compatibility
is often expressed in terms of database-derived properties and restraints.

In this review we analyze the attempt to use contact maps 
to efficiently perform protein fold prediction.
Contact maps are a particularly manageable representation of protein
structure which has been already applied in the past to the study
of protein conformation (Chan and Dill, 1990), 
structure comparison (Holm and Sander, 1993),
interaction patterns in proteins 
(Lifson and Sander, 1979; Godzik {\em et al.}, 1993)
and correlated mutations (Olmea and Valencia, 1997; Ortiz {\em et al.}, 1998).
The possibility of performing energy minimization in the space of contact
maps has been proposed by Mirny and Domany (1996).
We present here the consistent development of their idea, discussing
successes, failures and perspectives.

\section{Structure representation}
\begin{quote}
{\em We discuss the problem of representations of protein structure
and give the definition of contact maps.}
\end{quote}

Following Anfinsen's thermodynamic hypothesis (Anfinsen, 1973),
the native state of a protein is commonly assumed to be 
the minimum of a free energy function.
This is a powerful assumption and
Molecular Dynamics is the most direct method to implement it (Brooks, 1998).
The structure is represented by listing the
coordinates of all the atoms and Newton's equation of motion
are solved in a suitable force field tuned for molecular systems.
Unfortunately, present computers cannot follow the trajectory
of a protein all the way down to its native state. The best result
to date is the simulation of 1 $\mu$s trajectory of villin headpiece
subdomain in water which allowed the detection of the hydrophobic collapse
and of the formation of secondary structures 
(Duan and Kollman, 1998).
Furthermore, the method depends crucially on the determination 
of suitable energy parameters. An incorrect energy function,
which does not assign the lowest energy to the native conformation, leads a 
careful energy minimization procedure to some misfolded conformation,
as for example shown by Karplus and collaborators in the case of 
hemerythrin and the immunoglobulin VL domain 
(Novotny {\em et al.}, 1984).

We believe that perhaps one can solve the problem without
going to such microscopic detail
and we set about to investigate this assumption.
The problem of structure representation is to find
the best trade-off between computability and accuracy of the predictions.
Our inclination is that good predictions can be obtained
by constructing simplified models.

Lattice models offer the possibility to gradually turning on 
the complexity of the representation of the structure.
Usually a protein is represented as a chain of monomers occupying lattice
sites and representing $C_\alpha$ atoms.
The complexity can be measured by the number of states available
to each monomer (Park and Levitt, 1995). 
The lowest possible complexity is that of 
tetrahedral and simple cubic lattices, 
where 3 and 5 states per monomer are respectively available.
Lattices of high coordination number, up to 55 states, have been
studied (Ortiz {\em et al.}, 1998).
The main motivation to study lattice models
is that at low complexity it is possible to effectively solve
the problem of searching the ground state, either exactly, by enumerating
all the conformations or approximately, by Monte Carlo methods.
By using Monte Carlo simulations, solutions can be routinely obtained
for polymers up to length 125 
(\u{S}ali {\em et al.}, 1994; Dinner {\em et al.}, 1996).
Using pairwise contact energy functions, evidence for important features
of the folding process has been produced, most notably, 
the hydrophobic collapse.
It is possible to consider more detailed lattice models
which still retain some of the advantage in computability
and allow a more realistic representation of protein structure: for example,
it is possible to represent side chains by an additional virtual atom.
Advances in this direction have been recently reported by Skolnick and 
collaborators (Ortiz {\em et al.}, 1998).
Interestingly, the accuracy increases very slowly with the complexity,
and the typical resolution of structures in the Protein Data Bank (PDB) 
(Bernstein {\em et al.}, 1977) can be obtained with
models with 10 to 20 states (Park and Levitt, 1995).
The $C_\alpha$ model can be simulated also off-lattice.
For short chains and simple interactions it is possible to identify
the ground state with reasonable reliability
(Clementi {\em et al.}, 1998; Irb\"{a}ck and Potthast, 1995).

A minimalistic representation of a protein's structure is given by its
{\em contact map} 
(Lifson and Sander, 1979; Chan and Dill, 1990; Godzik {\em et al.}, 1993;
Hold and Sander, 1993; Mirny and Domany, 1996; Vendruscolo {\em et al.}, 1997).
The contact map of a protein with $N$ residues is an $N\times N$ matrix
${\bf S}$, whose elements are defined as
\begin{equation}
S_{ij}=
\left\{
\begin{array}{ll}
1 \qquad  & \mbox{\rm if residues $i$ and $j$
are in contact} \\
0 & {\rm otherwise}
\end{array} \right.
\end{equation}
One can define contact between two residues in different ways; one is to
consider two amino acids in contact when their two
$C_{\alpha}$ atoms are closer 
than some threshold $R_c$ (``$C_\alpha$'' definition) 
(Vendruscolo {\em et al.}, 1997).
Another definition is based on the minimal distance between 
two atoms that belong to the two residues 
(Hinds and Levitt, 1994; Mirny and Domany, 1996)
(``all atoms'' definition). In a contact map
$\alpha$-helices appear as thick bands of contacts
along the diagonal, $\beta$-sheets as bands running 
parallel or perpendicular to the diagonal.
Given all the inter-residue contacts or even a subset of them,
it is possible to reconstruct quite well a protein's structure, by means of
either Distance Geometry 
(Crippen and Havel, 1988), Molecular Dynamics 
(Br\"{u}nger {\em et al.}, 1986)
or Monte Carlo (Vendruscolo {\em et al.}, 1997). 

In contrast to Cartesian coordinates, the map representation of protein
structure is independent of the coordinate frame. This property made contact 
maps attractive for protein structure comparisons and for
{\it searching a limited database}  for similar structures 
(Chan and Dill, 1990; Godzik {\em et al.}, 1993; Hold and Sander, 1993).

One of the main reasons for selecting the contact map representation of
structure is our expectation and hope that we may be able to search the
space of contact maps in an efficient manner and find low energy maps (that
hopefully correspond to  conformations  close to the native one). 
In particular, one hopes that relatively simple changes on a map 
may generate very substantial coherent moves of the corresponding  
polypeptide chain conformation; moves which
would have taken much longer to achieve by working with the chain itself.

In order to actually implement such moves
in the space of contact maps, one has to overcome two important
problems. The main and foremost one is the need to ensure that the map 
${\bf S}$ that has been generated is {\it physical}.  
Our definition of ``physical'' will be given in detail below; 
broadly speaking, we mean that there exists a chain conformation whose
contact map is indeed our proposed ${\bf S}$. 
Arbitrary changes performed on a map yield, with very high probability,
non-physical maps. 
The reason is that the total number of possible $N \times N$ contact  
maps is $O(2^{aN^2})$, whereas the number of physical maps is much smaller, 
of order $O(2^{bN})$. 
The need for a procedure that limits the search to the small subspace
of physical maps was identified 
by Mirny and Domany (1996), who proposed a restricted set of moves.
The hope was that when moves selected from this set are performed
on a physical map, the new map is also physical. These rules were, however,
heuristic and no clear proof of the validity of this hope could be given.  

Subsequently, the problem of generating  physical maps was dealt with by
means of a {\it reconstruction procedure} (Vendruscolo {\em et al.}, 1997).
For any proposed contact 
map ${\bf S}$ one generates a chain, with ${\bf S}$  serving as the guide
of the construction process.
The procedure stops when the contact map ${\bf S}^\prime$ of the
resulting chain is close to the target map; this way one generates a
map which is physical {\it and}  close to the target.  

The second important issue is the
manner in which low energy maps are generated from an existing one 
(Vendruscolo and Domany, 1998a). 
The procedure has to be such that the resulting map is ``protein-like'', 
i.e. has secondary structure elements and the corresponding chain has 
the right density, bond-angle distribution,
chirality, etc. The final map should be physical and the decorrelation time
with the starting map should be short. 

In what follows we sketch how these two problems were addressed; for a more
detailed description of these procedures the reader is referred to the original
publications (Vendruscolo {\em et al.}, 1997; Vendruscolo and Domany, 1998a).

\section{The reconstruction procedure}
\label{sec:reconstruction}
\begin{quote}
{\em 
We present a method to obtain a three dimensional polypeptide conformation
from a contact map. We also explain how to deal with the case of
non physical contact maps.
}
\end{quote}

The aim is to find an efficient procedure, which can be performed ``on
line'' and in parallel with the dynamics in the space of contact maps, which
will ``project'' any map onto a nearby one which is guaranteed to be 
physically realizable. The protein is represented as a ``string of beads'',
in which each bead stands for an amino acid - the coordinates 
of center of a bead are identified with those of the corresponding $C_\alpha$ 
atom.  
For a given target contact
map \mbox{\bf S}, the algorithm  searches for a conformation 
that this string of beads
can take, such that the contact map \mbox{\bf S}$^{\prime }$ of our string
is similar (or close) to \mbox{\bf S}. If there exists a chain conformation
whose contact map is identical with \mbox{\bf S}, this contact map
is, by definition, physical. In general,
our method aims at converting a possibly
ill-defined, non-physical set of contacts to a legitimate one.
The three dimensional structure is in our case a means, rather than the end.

It should be noted that related, previously developed methods 
(Havel {\em et al.}, 1979; Br\"{u}nger {\em et al.}, 1986; 
Bohr {\em et al.}, 1993; Nilges, 1995; Lund {\em et al.}, 1996;
Aszodi and Taylor, 1996; Mumenthaler and Brown, 1996; 
Skolnick {\em et al.}, 1997)
a different aim; 
to construct three dimensional structures from measured distance information,
using various forms of distance geometry 
(Crippen and Havel, 1988),
supplemented by restricted Molecular Dynamics
(Scheek {\em et al.}, 1989)
or simulated annealing (Br\"{u}nger {\em et al.}, 1997).

One should emphasize here the
distinction between a contact map and a distance map.
In a contact map a minimal amount of information is available -
given a pair of amino acids, we know only if they are in contact or not.
That is, only lower and upper bounds on their separation are given.
A distance matrix, on the other hand, presents real-valued distances
between pairs of amino acids. 
The method presented here is not restricted to contact maps
and has been generalized to distance maps (Vendruscolo and Domany, unpublished).
The deviations between different structures 
that were reconstructed from the same contact map are typically much higher
than those between two structures derived from a distance matrix.

The proposed algorithm is divided into two parts. The first part, {\it growth,}
consists of adding one monomer at a time, i.e. a step by step growth of the
chain. The second part, {\it adaptation, } is a refinement of the structure,
obtained as a result of the growth stage, by local moves. In both stages, to
bias the dynamics, we introduce cost functions defined on the basis of the
contact map. Such cost functions contain only geometric constraints, and do
not resemble the true energetics of the polypeptide chain.

\subsection{Growth}

The first element of the growth is {\em single monomer addition.},
which is carried out in the spirit of the Rosenbluth method 
(Rosenbluth and Rosenbluth, 1955).
To add monomer $i$ to the growing chain 
we generate at random $N_{t}$ trial positions, (typically $N_t=10$),
\begin{equation}
{\bf r}_{i}^{(j)}={\bf r}_{i-1}+{\bf r}^{(j)}\;,
\end{equation}
where $j=1,\ldots ,N_{t}$. 
The length and the directions of ${\bf r}^{(j)}$ 
are set from a statistical analysis of PDB.
One out of the $N_{t}$ trials is chosen according to the probability 
\begin{equation}
p^{(j)}=\frac{e^{-E_{g}^{(j)}/T_{g}}}
             {\sum_{j=1}^{N_{t}}e^{-E_{g}^{(j)}/T_{g}}} \;,
\label{eq:prob}
\end{equation}
where $E_g$ is a cost function which
rewards contacts that should be present, according to the given contact
map, and discourage contacts that should not be there.

The second element of the procedure is {\em chain growth}.
The step by step growth just discussed optimizes the
position of successive amino acids along the sequence. The main difficulty
in the  method is that the single step of the growing chain has no
information on the contacts that should be realized many steps (or monomers)
ahead. To solve
this problem, we carry out several attempts (typically 10) 
to reconstruct the structure, 
choosing the best one. In practice, this is done as follows.

For each attempt, when position ${\bf r}^{[j(i)]}$ is chosen for monomer $i$
according to Eq. (\ref{eq:prob}), 
its probability is accumulated in the weight 
\begin{equation}
W_{i}=\prod_{k=1}^{i}p_{k}^{[j(k)]}
\label{eq:weight}
\end{equation}
When we have reached the end of the chain we store the weight $W_{N}$. The
trial chain with the highest $W_{N}$ is chosen.

The {\em cost function} for growth is
\begin{equation}
E_{g}{}^{(j)}=\sum_{k=1}^{i-1} 
d \cdot {\rm a}_g(S_{ik})\cdot \vartheta (d_{t}-r_{ik}^{(j)})\;,
\end{equation}
where $r_{ik}^{(j)}=|{\bf r}_{i}^{(j)}-{\bf r}_{k}|$.
The enhancing factor $d=i-k$ is introduced to guide the growth towards
contacts that are long ranged along the chain; $\vartheta $ is the Heaviside
step function and the constant ${\rm a}_g$ can take two values; 
${\rm a}_g(S_{ik}=0)\geq 0$ and ${\rm a}_g(S_{ik}=1)\leq 0$. 
That is, when a contact
is identified in the chain, i.e. $r_{ik}<d_{t}$, it is either rewarded
(when the target map has a contact between $i$ and $k$), or penalized. 

\subsection{Adaptation}

When we have grown the entire chain of $N$ points, we refine the structure
according to the following scheme. We choose a point $i$ at random and try,
using a crankshaft move (\u{S}ali {\em et al.}, 1994), 
to displace it to ${\bf r}_{i}^{\prime }$, keeping fixed the distances from
both points $i-1$ and $i+1$.
We use a local cost function $E_{a}^{(i)}:$ 
\begin{equation}
E_{a}^{(i)}=\sum_{k=1}^{N}
f_{a}(r_{ik}^\prime)={\rm a}_a(S_{ik})\cdot \vartheta (d_{t}-r_{ik}^\prime)\;,
\end{equation}
where $r_{ik}^{\prime }=|{\bf r}_{i}^{\prime }-{\bf r}_{k}|$.
Note that the enhancing factor $d$ has been omitted, so that $f_{a}$ 
does not favor contacts between monomers that are distant along the chain.
The displacement is accepted with probability $\pi $, according to the
standard Metropolis prescription 
\begin{equation}
\pi =\min (1,\exp (-\Delta E_{a}/T_{a})
\end{equation}
where $\Delta E_{a}$ is the change in the cost function $E_{a}$ induced by
the move and and $T_{a}$ is a temperature-like parameter, used to control
the acceptance ratio of the adaptation scheme.

\begin{figure}
\centerline{\psfig{figure=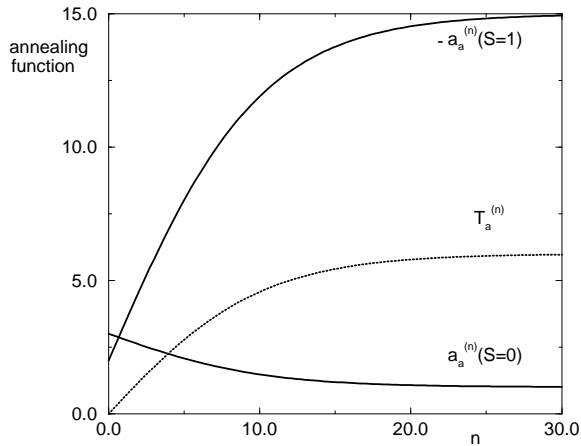,height=7.0cm,angle=0}}
\caption{Annealing functions for the parameters used in adaptation.
See Eqs. (\protect\ref{eq:a}) and (\protect\ref{eq:t}).}
\label{fig:annealing}
\end{figure}

A key ingredient of our method is annealing 
(Kirkpatrick {\em et al.}, 1983). As 
in all annealing procedures,
the temperature-like parameter $T_a$ is  decreased slowly during the
simulation to help the system find the ground state in a rugged energy
landscape.
In our method, however, 
instead of using simulation time as a control parameter on
the temperature, we chose the number $n$ of missing contacts. Two regimes were
roughly distinguished. In the first regime many contacts are missed and
the map is very different from the target one. In the second regime
few contacts are missed, and the map is close to the target. The
parameters ${\rm a}_a$ and $T_{a}$ are  interpolated smoothly between 
values suitable for these two limiting
cases. In the first regime, we strongly favor the recovery of contacts that
should be realized, whereas in the second regime we strongly disfavor contacts
that are realized but should not be present. We set, 
as shown in Fig. \ref {fig:annealing}
\begin{equation}
{\rm a}_a^{(n)}(S)={\rm a}^{f}(S)+[{\rm a}^{i}(S)-{\rm a}^{f}(S)]\sigma (n)\;,
\label{eq:a}
\end{equation}
and
\begin{equation}
T_{a}^{(n)}=T_{a}^{f}+(T_{a}^{i}-T_{a}^{f})\sigma (n)\;.
\label{eq:t}
\end{equation}
The function $\sigma (n)$ interpolates between the initial 
value ${\rm a}^{i}$ and the final value ${\rm a}^{f}$, 
\begin{equation}
\sigma (n)=\frac{2}{1+e^{-\alpha_g n}}-1\;.
\end{equation}
By choosing ${\rm a}^{i}$, ${\rm a}^{f}$, 
$T_{a}^{i}$, $T_{a}^{f}$ and $\alpha_g $ we
define the two regimes, far from  and close to the target map. 

We have tried several alternatives to each of the components of the method
outlined above. For a detailed description of these, we refer the reader to 
Vendruscolo {\em et al.} (1997).
We present here a brief description of some selected results 
that were obtained using the algorithm presented above.

\subsection{Results}
\subsubsection{Experimental contact maps.}

In this section we present results about the reconstruction of experimental
contact maps as taken from PDB. Since our purpose,
as explained in the Introduction, is to use the
reconstruction in connection with dynamics, 
we chose the contact length $R_c=9$ \AA \hspace{3pt} to
obtain the most faithful representation of the energy of the protein 
(Mirny and Domany, 1996).
Such a threshold is determined by the requirement 
that the average
number of $C_{\alpha}-C_{\alpha}$ contacts for each amino acid is roughly
equal to the respective numbers obtained with the all-atom definition of
contacts.

The most commonly used dissimilarity measure
between structures is
the root mean square (RMS) distance $D$ (McLachlan, 1979) 
\begin{equation}
D = \sqrt{\frac{1}{N} \sum_{i=1}^N ( {\bf r}_i - {\bf r}_i^{\prime})^2} \; ,
\label{eq:distance}
\end{equation}
where one structure is translated and rotated to get a minimal $D$.

The dissimilarity measure between contact maps is defined as 
the Hamming distance
\begin{equation}
D^{{\rm map}} = \sum_{j>i} | S_{ij} - S_{ij}^{\prime} | \;,
\end{equation}
which counts the number of mismatches between maps 
${\bf S}$ and ${\bf S^{\prime}}$.

\begin{figure}
\centerline{\psfig{figure=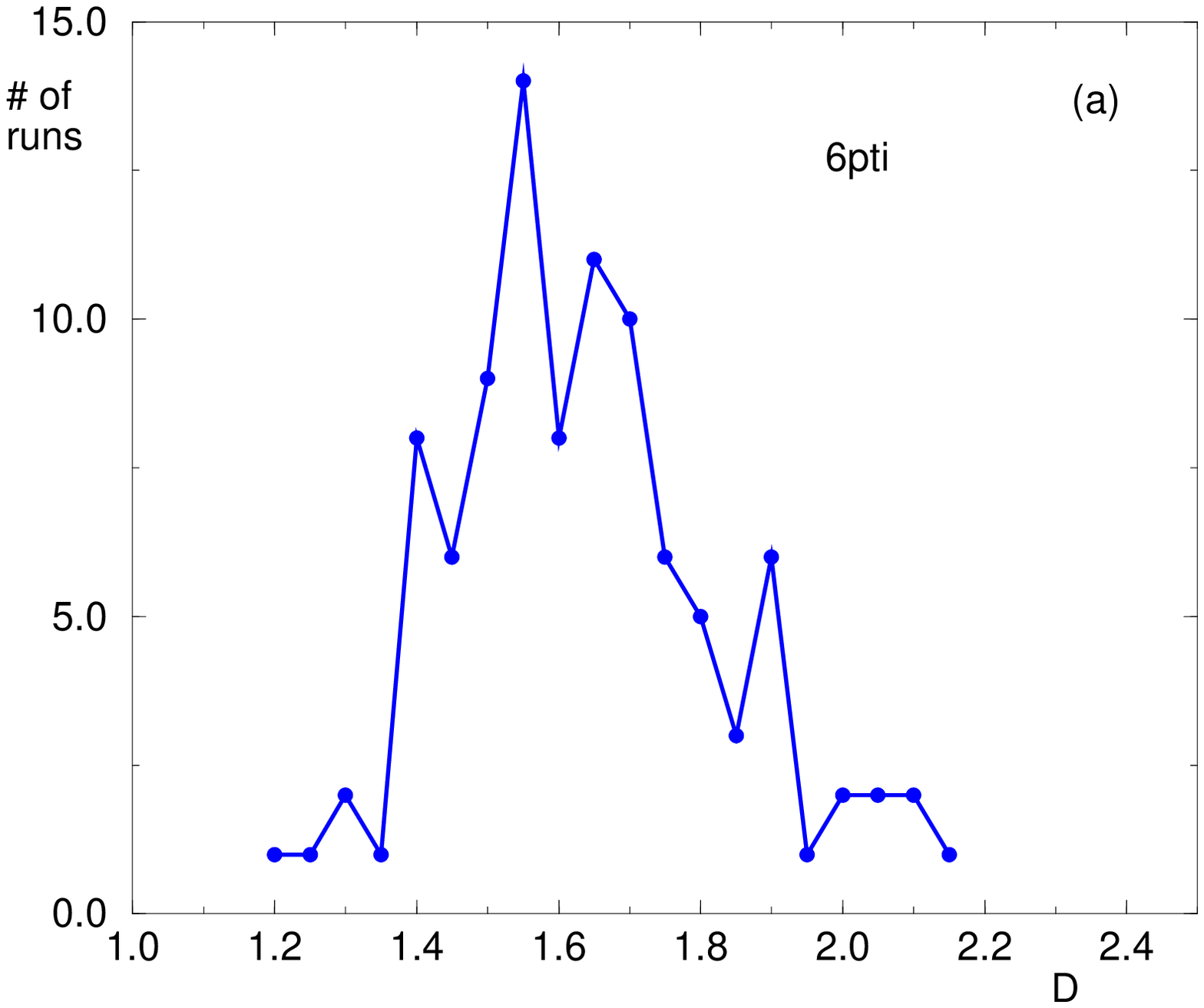,height=7.0cm,angle=0}
            \psfig{figure=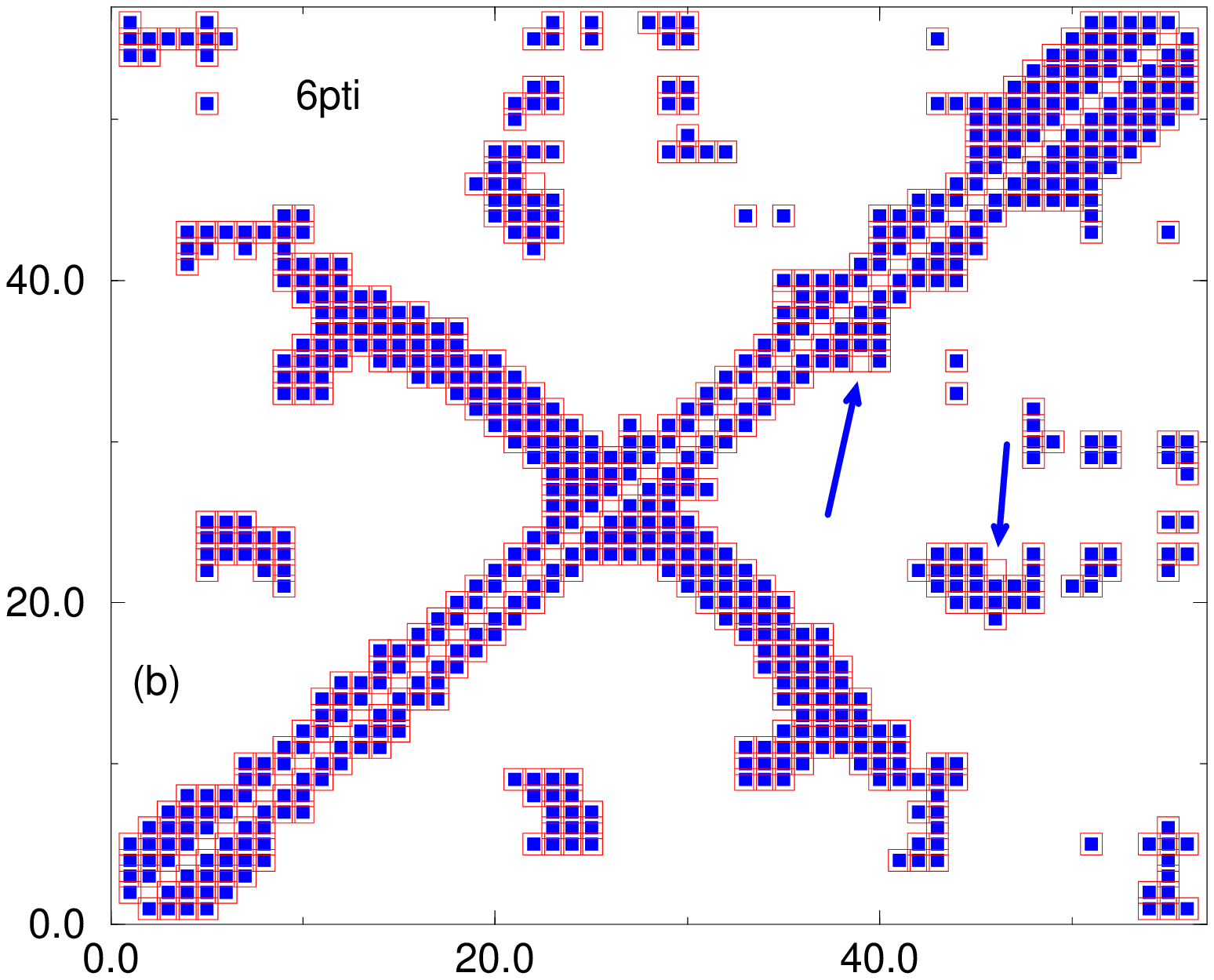,height=7.0cm,angle=0}}
\caption{(a) Histogram of the distance $D$ 
for the 100 runs used to test the reconstruction
procedure. Data are presented for protein 6pti.
(b) Contact maps for protein 6pti for a threshold $d_t = 9$ \AA.
Full squares are the PDB data,
open squares the output of the reconstruction procedure.
None of the target contacts is missed and two spurious ones are added
(the arrows point at their locations.
(Adapted with permission from Vendruscolo {\em et al.} (1997).
Copyright 1997 Current Biology.)}
\label{fig:runs}
\end{figure}

We present in  Fig. \ref{fig:runs}a the results
over 100 reconstruction runs for chains 
for the protein 6pti (bovine pancreatic trypsin inhibitor).
In Fig. \ref{fig:runs}b we show the contact map for the protein 6pti,
$N=56$, as taken from PDB, that was used as a target to construct
a chain. The contact map of a typical reconstructed chain is also shown.
In this particular case none of the 342 original contacts were missed and only
two false positive contacts were added. These are close to clusters of correct
contacts, indicating slight local differences with the crystallographic 
structure.
The distance recorded in this case was $D=1.56$. 

We carried out extensive similar tests for other proteins of various lengths
(Vendruscolo {\em et al.}, 1997).
Our method produces, using
a native contact map as target, a structure whose contact map is in nearly
perfect agreement with the target. 
Furthermore, the distance of this reconstructed chain  from the native structure
is quite close to the resolution that can be 
obtained from the information contained in contact maps.

\subsubsection{Non-physical contact maps}
Our main purpose is to develop a strategy to construct a three dimensional
structure, starting from a given set of contacts, even if these
contacts are not physical, i.e. not compatible with any 
conformation allowed by a chain's topology.	 In such a case we require 
our procedure to yield a chain whose conformation is 
as ``close'' as possible to the contact map
we started with. The exact measure of such closeness 
depends on the source of non-physicality, as will be demonstrated in 
two examples described below.

Our first examples of non-physical contact maps 
were obtained by randomizing a native contact
map; this was done by flipping $M$ randomly chosen entries.
Contacts between consecutive amino acids (neighbors along the chain) 
were conserved.

\begin{figure}
\centerline{\psfig{figure=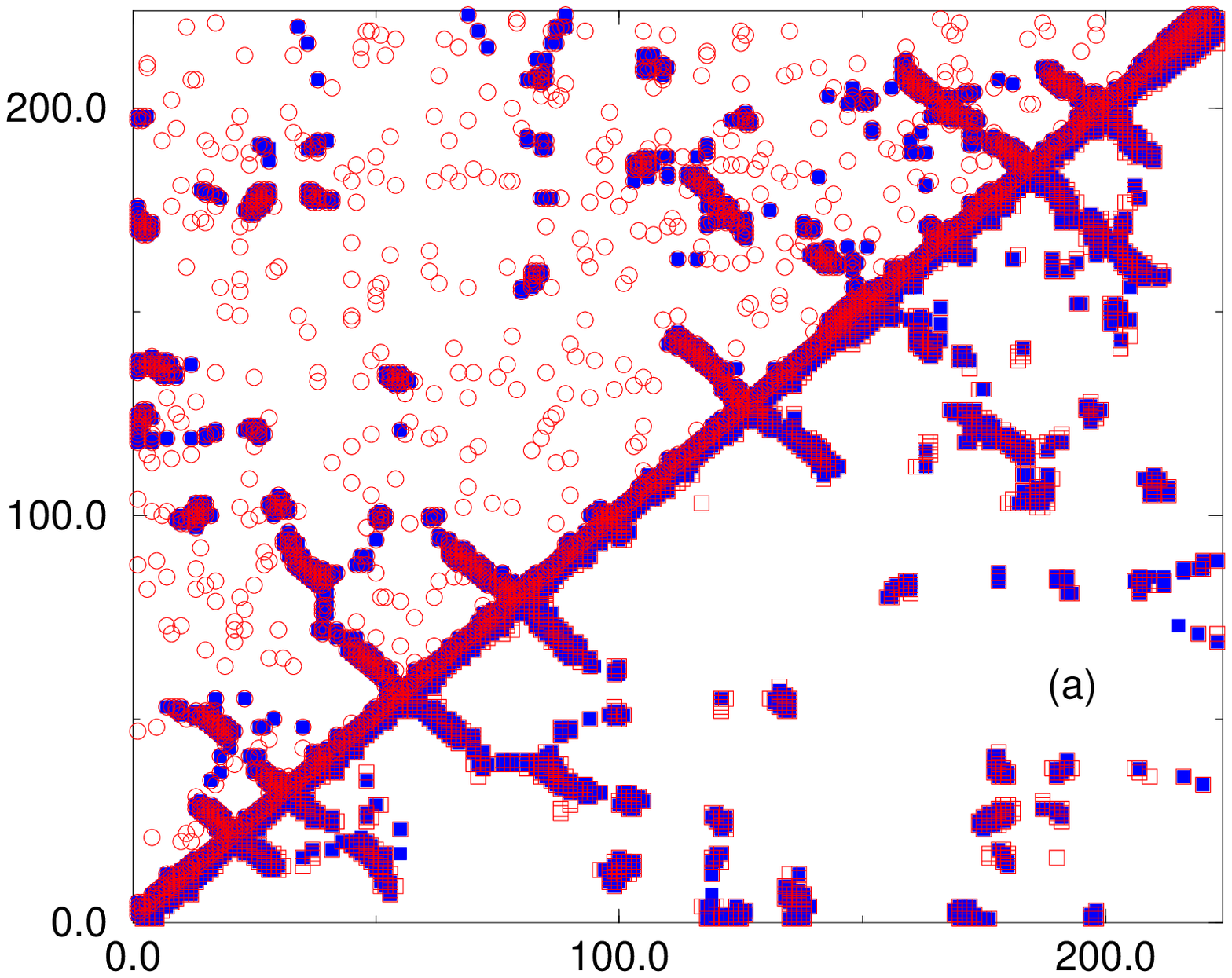,height=7.0cm,angle=0}
            \psfig{figure=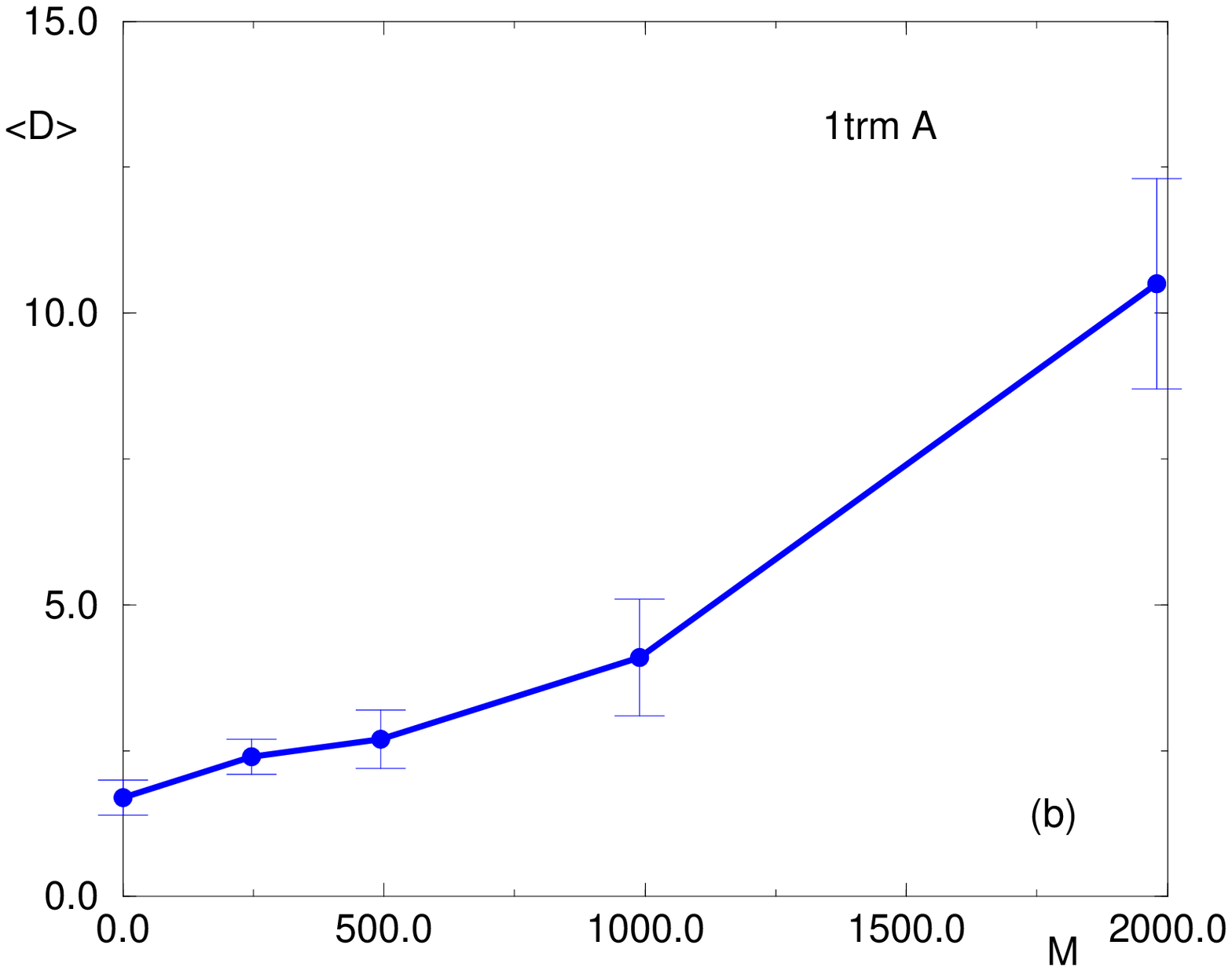,height=7.0cm,angle=0}}
\caption{
(a) Above diagonal:   reference map
(open circles) obtained by randomizing the 
underlying physical map (full squares)
of protein 1trm chain A.  
Below diagonal: reconstructed contact map (open square)
obtained using the noise-corrupted map as target.
(b) Average distances $\langle D \rangle$ 
versus noise $M$ for protein 1trm A.
(Adapted with permission from Vendruscolo {\em et al.} (1997).
Copyright 1997 Current Biology.)}
\label{fig:1trma_dis_noise}
\end{figure}

A typical contact map with noise is shown in Fig. \ref{fig:1trma_dis_noise}. 
The protein is 1trm chain A, 
whose contact map has 1595 contacts, when the threshold is set to 9 \AA.
We show the native map and the target
map obtained by flipping at random $M=400$ 
entries of the native map, together with 
the map  produced by our method.
For the particular case shown, the distance to the crystallographic
structure $D=2.4$ \AA. 
The most important conclusion that can be drawn from 
Fig. \ref{fig:1trma_dis_noise}
is that isolated non-physical contacts are identified as such  and ignored
and the underlying physical contact map is recovered.

The dependence of this recovery on the noise level is
shown in Fig. \ref{fig:1trma_dis_noise}, where we present
the average distance of 
the final structure from the uncorrupted 1trm A contact map, 
for various values of $M$.
Averages were taken over 10 different realizations of the noise, and
over 10 reconstruction runs for each realization.
The distance for totally unrelated structures 
for 1trm A is around 15 \AA. 
It is remarkable that up to $M<1000$ a fair resemblance to the experimental
structure is still found. 
Even with the addition of a noise which is around 60\% of the signal,
the reconstruction procedure works.
We have found similar result for the smaller protein 6pti, which has
342 contacts, and can be fairly well 
reconstructed with a noise of up to 200 flipped contacts.

To summarize this section, 
for physically realizable target contact maps  
our method is very fast and reliable to find a chain conformation 
whose contact map is nearly identical to the target.
Moreover, a the method is able to find a good candidate structure even when
the target map has been corrupted with non-physical contacts.
The information contained in a known native contact map  
suffices to reconstruct
a conformation, which is relatively close to that of the original structure, 
as was already observed by Havel {\em et al} (1979).
There is, however, an intrinsic limit in the resolution of a contact map.
We used  a threshold of $9$ \AA \hspace{3pt}
between $C_\alpha$ atoms to define contact; for this threshold
the distance between two typical structures, that are both
compatible with the contact map, is about $1$ \AA. The threshold of
$9$ \AA \hspace{3pt} is relevant for our purpose, of 
working with contact energies in
a scheme to derive structure from sequence. 


\section{Dynamics in contact map space}
\label{sec:dynamics}
\begin{quote}
{\em 
We describe a stochastic method to perform dynamics in contact map space.
We explain how the motion is restricted to physical regions of the space.
}
\end{quote}

Our aim is to generate a large number of contact maps that can serve as
candidates for the native structure. Such maps are necessary for protein
folding by means of energy minimization, as well as in order to generate
decoys needed to test properties of various energy functions. Hence the
requirements from any procedure that generates such maps are

\begin{itemize}
\item
The generated maps should be physical.
\item
The maps should be ``protein-like"; for example they should
have secondary structure elements
\item
The maps should have low values of the energy (defined in terms of
the sequence and the contact map)
\item
Efficiency - in order to generate large numbers of independent maps
in reasonable computing time.
\end{itemize}

The requirement of physicality is addressed by the method 
described in the Sec. \ref{sec:reconstruction}; 
whenever a new candidate map is generated, 
we use it as the target map of the reconstruction procedure, 
and obtain, this way, a contact map which corresponds to
a physical ``chain of beads''.

In order to move efficiently in contact map space in a way that
satisfies the requirements listed above, we introduced a
four-step procedure which is delined below. For further details
we refer to Vendruscolo and Domany (1998a).

\begin{enumerate}
\item
Non-Local Dynamics: \\
Starting from an existing map, we perform large scale ``cluster'' moves. 
Clusters are in approximate correspondence
with secondary structure elements.
At this stage, no attempt is made to preserve physicality.
The contact map which is obtained by this procedure is typically
uncorrelated to the starting one.
\item
Local Dynamics: \\
The resulting map is refined by using local moves of different kinds.
Secondary structure formation can be viewed as a ``growth'' process.
Starting from a random coil, an $\alpha$ helix is formed by twisting
one turn at a time (Chakrabartty and Baldwin, 1995).
Analogously, an helix can translate locally
by untwisting a turn at one extreme and reforming it at the opposite end,
in a movement reminiscent to the reptation dynamics of polymers 
(de Gennes, 1979).
A $\beta$ sheet is created and removed by zipping and unzipping together
two $\beta$ strands (Mu\~noz {\em et al.}, 1997).
We also use the conservative dynamics introduced by Mirny and Domany (1996)
to further refine the resulting map.
\item
Reconstruction: \\
We use the previously introduced reconstruction algorithm 
(Vendruscolo {\em et al.}, 1997)
to restore physicality by projecting the map obtained from the second step
onto the physical subspace.
\item
Refinement: \\
We perform  further optimization
by  energy minimization
in {\it real space} using  standard crankshaft moves 
(\u{S}ali {\em et al.}, 1994; Vendruscolo and Domany, 1998a).
\end{enumerate}

The projection procedure from a contact map to its three dimensional
counterpart is the bottleneck of the method. The dynamic rules
that we introduce are aimed at generating uncorrelated starting points
for this reconstruction. In this way, after each four-step move,
we obtain a good candidate map for the native state.
The contact energy (eq. \ref{eq:pair})
with some standard parametrization (i.e. choice of the $w(a_,b)$)
is used in steps 1, 2 and 4
following the standard Metropolis prescription for the acceptance
of a move.

\subsection{Non-local dynamics}

Rules of non-local dynamics have been introduced in the context of the
simulation of equilibrium properties of spin systems 
(Kandel and Domany, 1991)
and of surfaces (Evertz {\em et al.}, 1991).
Under suitable conditions,
systems with a large number of degrees of freedom
arrange themselves in conformations where the degrees of freedom
are ``coherently'' grouped together.
Using an incoherent dynamic procedure the time it takes to go from
one coherent conformation to another can be prohibitively long.
Physical intuition guides the choice of non-local rules
to obtain an efficient dynamics.
Since in our case we are developing a minimization algorithm,
we are not concerned with detailed balance, and we can 
optimize the dynamics by choosing moves that minimize the energy.

We now present the physical considerations that guided our
choice of non-local moves.
The unknown interactions between amino acids dictate the rules
that determine the stability of protein folds. 
Such rules govern the chain topology in a rather stringent way. 
The overall number of existing protein families
is estimated to be around 1000 
(Chothia, 1992; Orengo {\em et al.}, 1994).
Protein families are characterized by particular arrangements of
secondary structures. Secondary structure elements 
are easily identified also in a contact map
as clusters of points of characteristic size, shape and location.
In the contact map representation, 
secondary structures can be handled very efficiently by binary operations.
For example a parallel ${\beta}$ sheet is created by turning from 0 to 1
a set of contacts forming a cluster with the shape of a thin band 
parallel to the main diagonal.
To turn a 2-${\alpha}$ bundle
from an up-down topology to the alternative up-up one,
a rotation of a cluster of points is required.
The operations described provide only a scaffold, which is non physical, and
must be rectified by the other two steps of the dynamics.
The procedure yields every time a completely new topology.
A MD simulation could obtain the same result only by 
completely (or at least partially) unfolding and refolding the protein.

In contact maps of experimentally determined
protein structures clusters of contacts
can be divided in four classes. 
Thick bands of adjacent contacts along the main diagonal 
represent $\alpha$ helices (see region 1 in Fig. \ref{fig:1ubq}a). 
Thin bands represent parallel $\beta$ sheets 
if they are parallel to the main diagonal 
(region 2 in Fig. \ref{fig:1ubq}a)
and antiparallel $\beta$ sheets if they are antiparallel
(region 3 in Fig. \ref{fig:1ubq}a).
Small clusters or isolated points represent
structurally relevant contacts between amino acids that
are well separated along the sequence.
These features characterize protein-like contact maps and
should be preserved by the dynamics in contact map space.

As a preliminary information we determine the expected number
$N_c^*$ of contacts and the number $N_s^*$ of clusters
that are expected to appear in the contact map.
These numbers will be stochastically conserved
during the dynamics.
We have already presented evidence 
to the effect (Vendruscolo {\em et al.}, 1997) that 
$N_c^* = a N^{\nu}$, where $N$ is the length of the protein,
$\nu \simeq 1$ and $a$ depends on the threshold that is used
to define a contact.
As for $N_s^*$, there are algorithms 
to predict the secondary structure content, like the PHD 
(Rost and Sander, 1993) 
or the GOR (Garnier {\em et al.}, 1996) algorithms.
Alternatively, having a good starting guess for the native contact map,
one can directly determine $N_c^*$ and $N_s^*$.
The cluster algorithm is divided into three steps:
labeling, deletion and creation.

\begin{enumerate}
\item
{\em Labeling:}\\
Starting from an existing map, the first step 
is to identify the clusters that are present, which is done using
the Hoshel-Kopelman algorithm 
(Stauffer and Aharony, 1992).
With our definition of contact,
contacts $S_{ij}$ with $|i-j|\le 2$ 
are always present due to chain connectivity,
see Fig. \ref{fig:1ubq}a.
Our dynamical rules do not violate these topological constraints.
Cluster labeling is made in the upper triangle, excluding the first three
diagonals. By symmetry, it is sufficient to perform all the dynamics
in this region.
At this stage we calculate the number $N_c$ of contacts and the number $N_s$
of secondary structures. Secondary structures are defined as cluster
of more than $H=10$ points. 
After this labeling procedure, each point $(i,j)$ in the map has label
$L(i,j)=(C,K)$ with a class $C$ and a number $K$ inside the class.
Five classes are considered. In class $\alpha$
we put the clusters that are formed by bands along the main diagonal. In
class $\beta_\parallel$ we put the clusters that constitute
bands parallel to the main
diagonal but apart from it. In class $\beta_\perp$ we group
clusters that are in the form of bands perpendicular
to the main diagonal. In the fourth class
we gather all the remaining irregular clusters. In the last class
we put all the points which do not belong to any cluster
(e.g. isolated points).

\item
{\em Destruction:}\\
$N^-$ existing clusters are deleted from the map.
$N^-$ is chosen from a uniform random distribution between 1 and $N_s$.
Destruction is simply realized by choosing at random a label $(C,K)$
and by turning contacts in the corresponding cluster from 1 to 0.

\item
{\em Creation:}\\
$N^+$ clusters are created in the map.
$N^+$ is drawn from a gaussian distribution of mean $N_s^*-(N_s-N^-)$
and variance 1. If $N_s-N^- > N_s^*$ then $N^+=0$.
Each time we make $M$ attempts to create a cluster (typically $M=100$), 
and we choose the one with
the more favorable energy, according to Eq. (\ref{eq:pair}).
At each creation we first decide with probability $P(\alpha)$, 
$P(\beta_\parallel)$ and $P(\beta_\perp)$ 
whether to grow an $\alpha$, a $\beta_\parallel$ or a $\beta_\perp$.
Typically $P(\alpha)=P(\beta_\parallel)=P(\beta_\perp)=1/3$.
The length of the created band
is a uniform random number in [5,30] for $\alpha$,
in [5,12] for $\beta_\parallel$ and $\beta_\perp$.
Creation starts by selecting randomly a seed point on the map.
For $\alpha$ clusters this point is chosen on the principal diagonal,
for $\beta_\parallel$ at a point displaced from the principal diagonal by more
than the length of the cluster. No restrictions are imposed 
on the seed of $\beta_\perp$.
From this point we lay down a cluster 
in the form of a band as shown in Fig. \ref{fig:1ubq}b.
We do not allow secondary structures to overlap or to be closer than 4
spacings on the map, 
since this is not commonly observed in actual contact maps.
If while growing the cluster we encounter a point
which already has a label that violates this condition
we restart the creation.
The result of a non-local move for 1ubq, starting from the native map
shown in Fig. \ref{fig:1ubq}a is shown in Fig. \ref{fig:1ubq}b.

\end{enumerate}

\subsection{Local dynamics}

The principal aim of these moves is to allow local rearrangements
of the secondary structures that have been placed by the non-local dynamics.
We first give the dynamical rules
to deal with $\alpha$ helices. Consider an helix of $n$ amino acids, 
which we have previously identified as 
starting from amino acid $i$ and ending on $i+n$. 
Typically, $n$ ranges from 5 to 30.
To increase the size from the head, we add the two contacts
$(i+n+1,i+n-2)$,  $(i+n+1,i+n-3)$.
The tail is increased by adding the two contacts 
above the diagonal $(i-1,i+2)$ and $(i-1,i+3)$.
To decrease the size of the helix, one removes the contacts
$(i+n,i+n-3)$ and $(i+n,i+n-4)$ on the head and
$(i,i+3)$ and $(i,i+4)$ on  the tail.
To translate the helix, one performs a reptation-like move 
in which one turn is removed from one end and added to the other,
by using the same rules.

Similar rules govern the growth and the translation of sheets.
Consider first an antiparallel $\beta$ sheet formed by two strands.
The first strand extends from amino acids $i$ to $i+n$ and the second
from $j$ to $j+m$.
By unzipping amino acids $i+n$ and $j$, we
reduce the size at the end closer to the main diagonal.
This move is realized by setting to 0 (irrespective of their state)
the five contacts $(i+n,j+2)$, $(i+n-1,j+2)$, $(i+n-1,j+1)$, $(i+n-2,j+1)$
and $(i+n-2,j)$.
Opening the sheet from the other side is realized by setting to 0
the contacts $(i+2,j+m)$, $(i+2,j+m-1)$, $(i+1,j+m-1)$,
$(i+1,j+m-2)$ and $(i,j+m-2)$.
Zipping together the ends is realized by setting the corresponding
contacts to one.
Translating the sheets, as in the case of helices, is realized
by opening one end while closing the other.
Rules that are entirely similar are applied to parallel $\beta$ sheets.
In the general case, the sheet might present irregularities
which would appear as supplementary contacts at the extremities.
Since we do not attempt here to realize a physical map
we implement these simple rules and rely on the reconstruction procedure
to take care of the local structural details.
The resulting contact map after this step is shown in Fig. \ref{fig:1ubq}c.
We use the conservative dynamics introduced by Mirny and Domany (1996)
to further refine the resulting map;
the result is shown in Fig. \ref{fig:1ubq}d.
Typically minor local rearrangements take place.

\subsection{Reconstruction}
As already observed, a generic contact map 
is not guaranteed to correspond to any real chain conformation in space.
This is likely the case of the contact map obtained after the first 
two steps of the dynamics.
By using the reconstruction method presented above
we project this contact map onto its closest physical counterpart,
i.e., we create
a contact map which is ``close'' to the starting one, as measured 
by the Hamming distance,
and is guaranteed to be physical, i.e., there is a real chain
conformation which has that contact map.
To achieve this result, we construct a backbone conformation in 
cartesian space and try to force it to have the contacts specified 
in the input contact map. 
If an input contact map is non-physical, existing 
contacts are discarded and possibly new ones are introduced.
Since, however, any difference in the number and locations of contacts
is penalized, the contact map of the resulting conformation is necessarily
close to the starting one (Vendruscolo {\em et al.}, 1997).
Monomers are not allowed to invade each other's space. This is
ensured by introducing
a lower threshold $R_L$, below which they experience a hard-core repulsion.
The lowest $C_\alpha$-$C_\alpha$ distance found in PDB proteins
is around 3.5 \AA. We chose $R_L$=5.0 \AA.
With such a value, it is still possible to reconstruct all the PDB proteins
and the tendency to create too compact structures, typical of the
contact energy approximation, is minimized.
Result of the reconstruction is shown in Fig. \ref{fig:1ubq}e.

\subsection{Refinement:}
We perform further optimization by an energy minimization
in real space using a standard Metropolis crankshaft technique
(\u{S}ali {\em et al.}, 1994; Vendruscolo and Domany, 1998a).
Result of the minimization is shown in Fig. \ref{fig:1ubq}f.
In this calculation we used a set ${\bf w}_{153}$ of contact energy parameters,
that was derived using the method 
presented by Mirny and Domany (1996),
applied to the database of 153 proteins listed in Vendruscolo {\em et. al}
(1998) with the present definition of contact.
The initial energy of the native fold of 1ubq is 25.72 and the energy of the
final map is much lower, -84.20.

\begin{minipage}[t]{6.5in}
\begin{figure}
\centerline{\psfig{figure=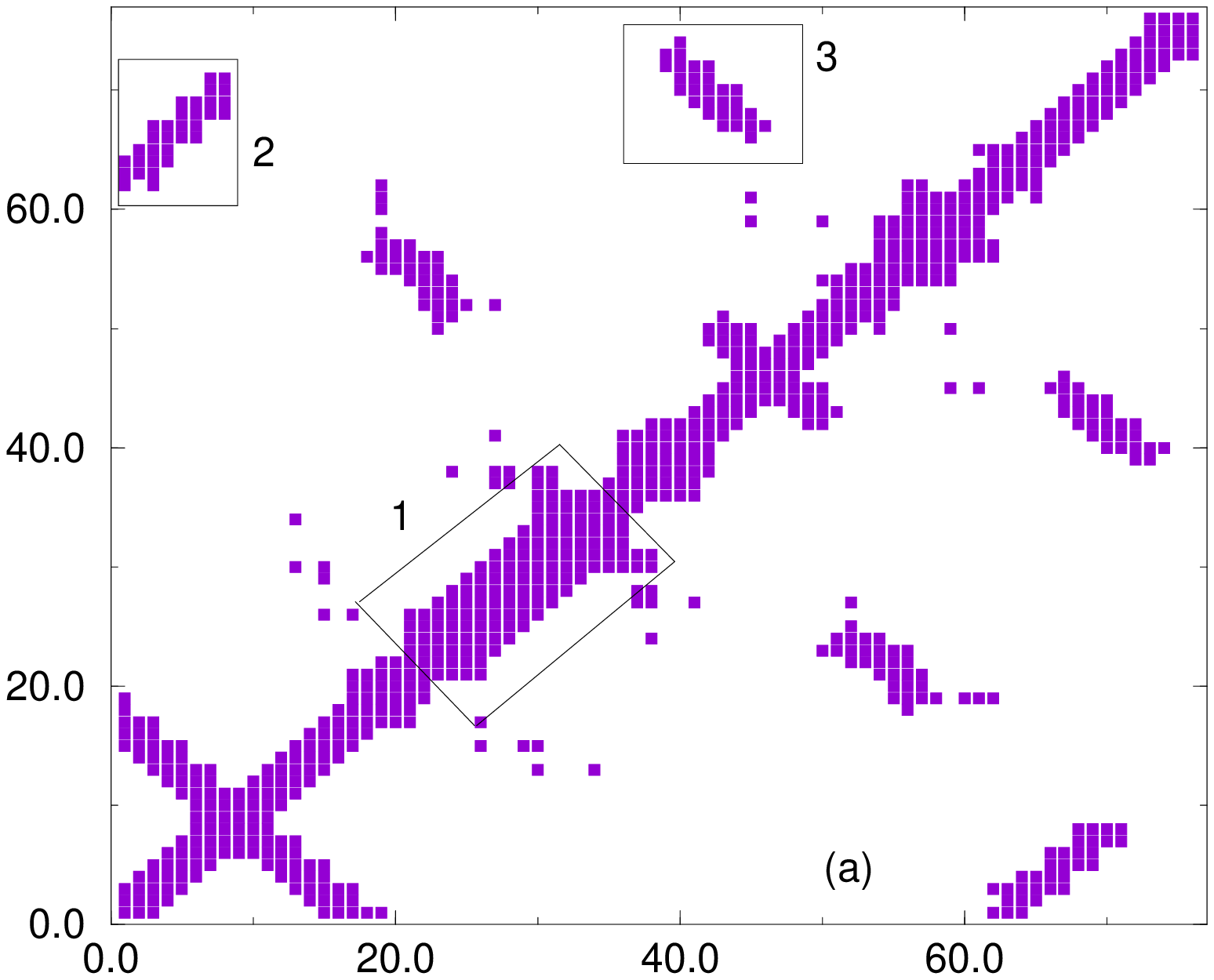,height=6.0cm,angle=0}
            \psfig{figure=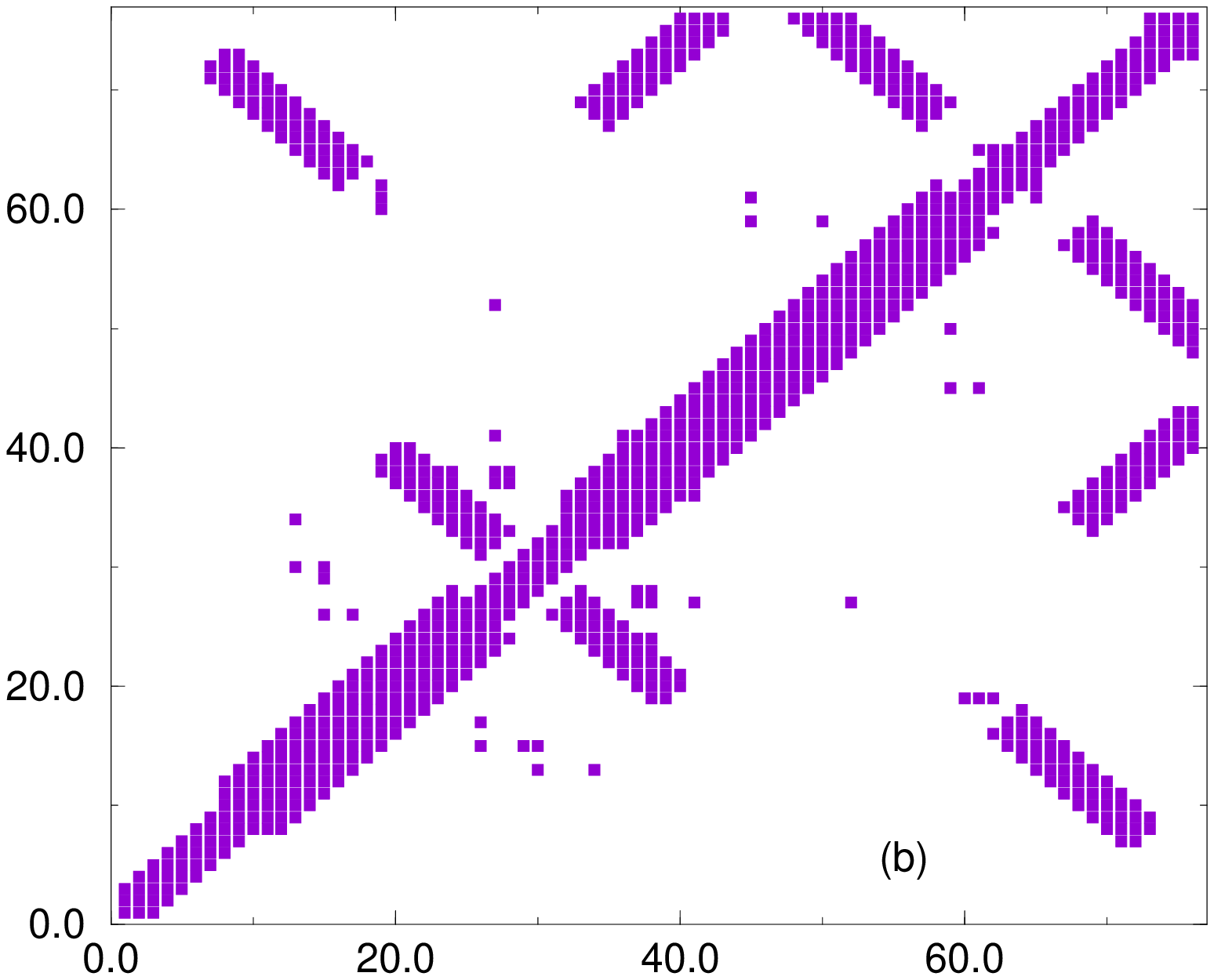,height=6.0cm,angle=0}}
\centerline{\psfig{figure=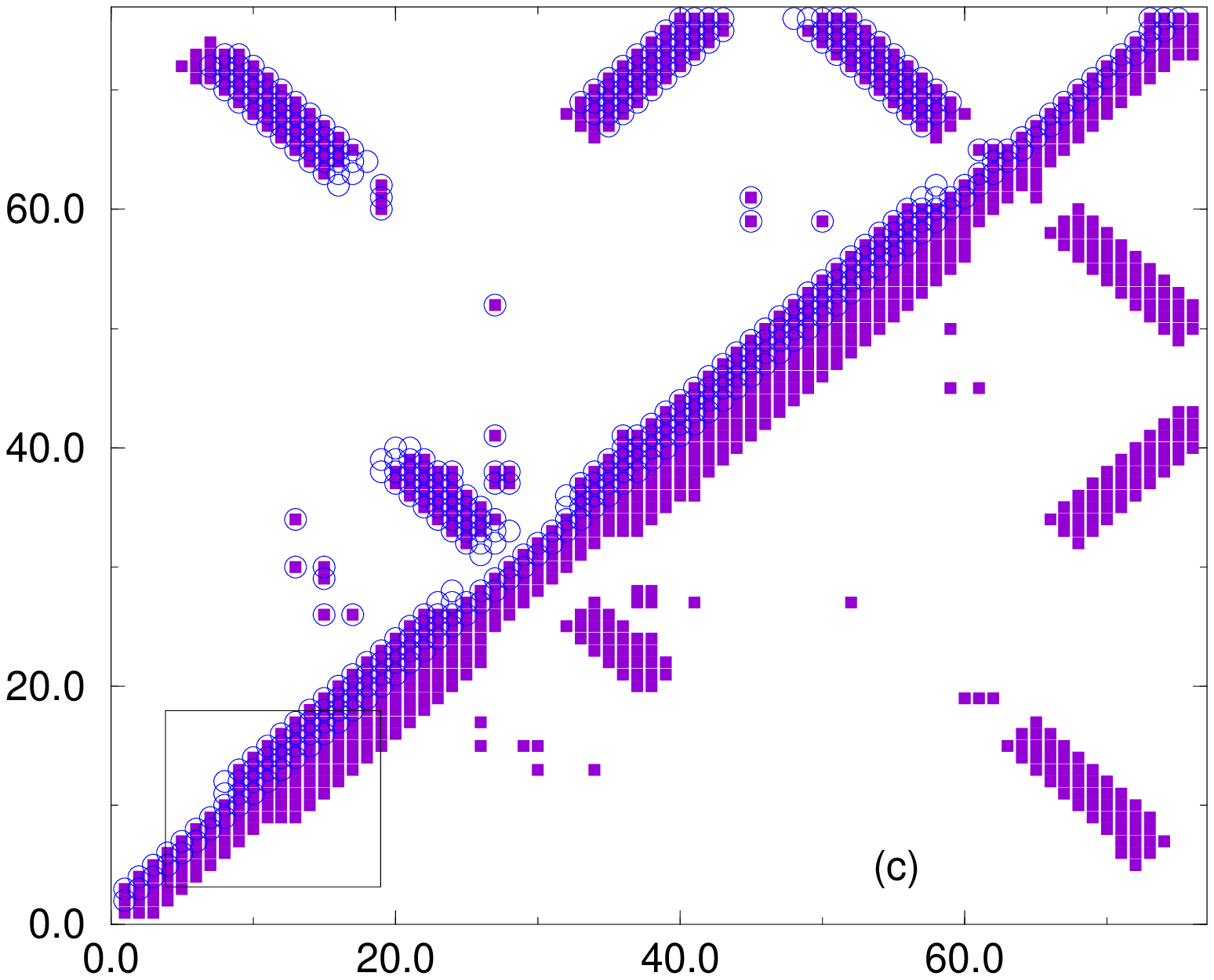,height=6.0cm,angle=0}
            \psfig{figure=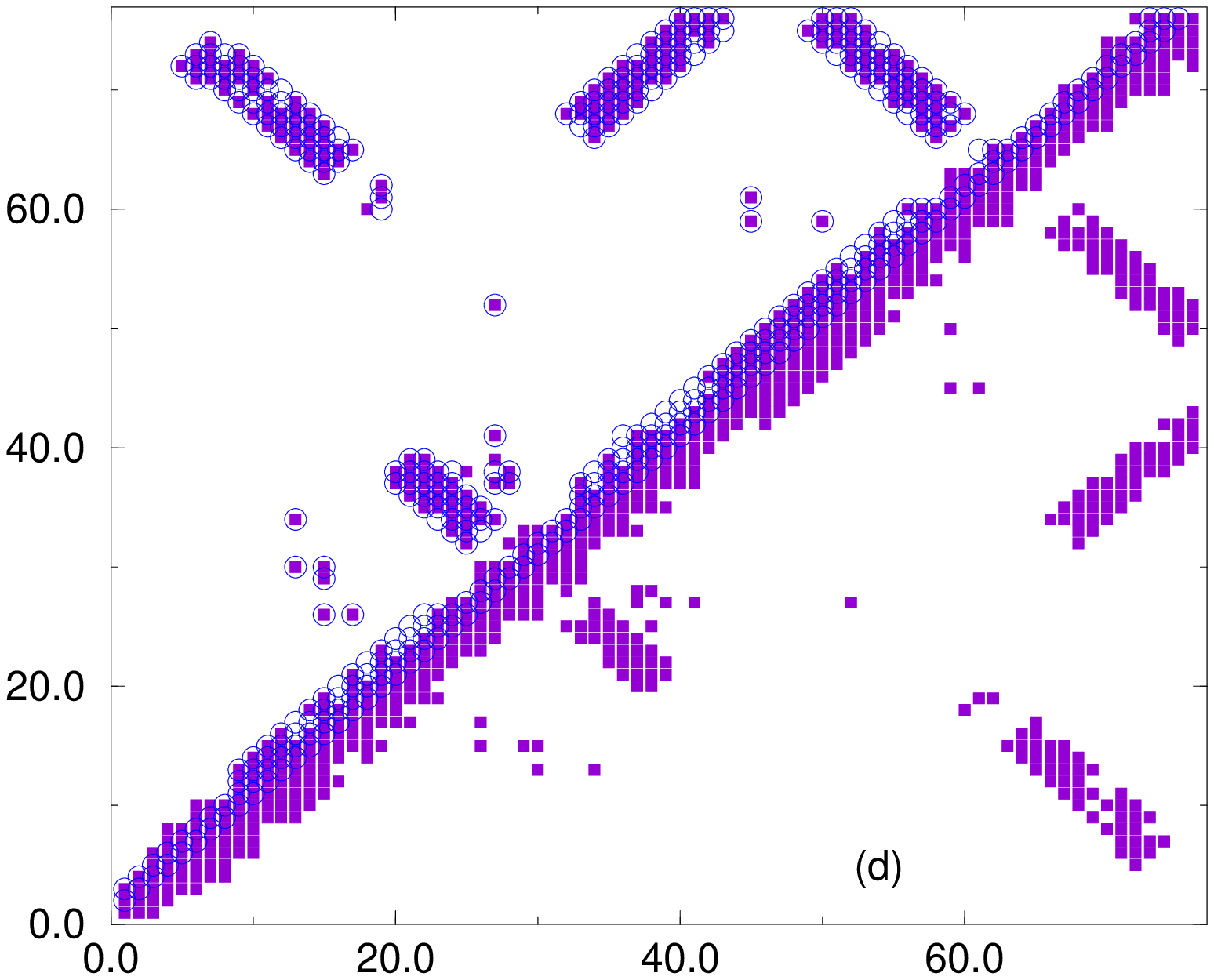,height=6.0cm,angle=0}}
\centerline{\psfig{figure=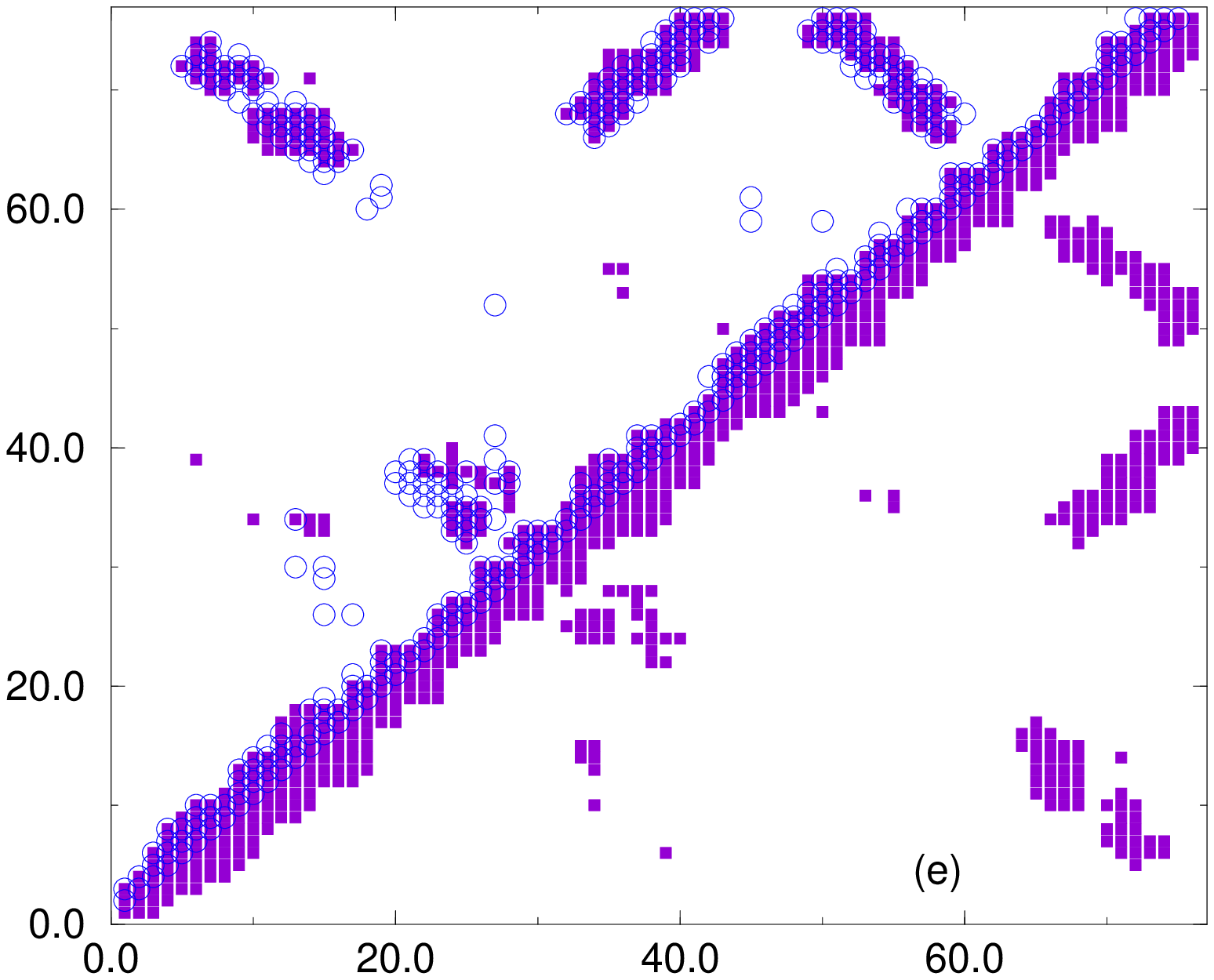,height=6.0cm,angle=0}
            \psfig{figure=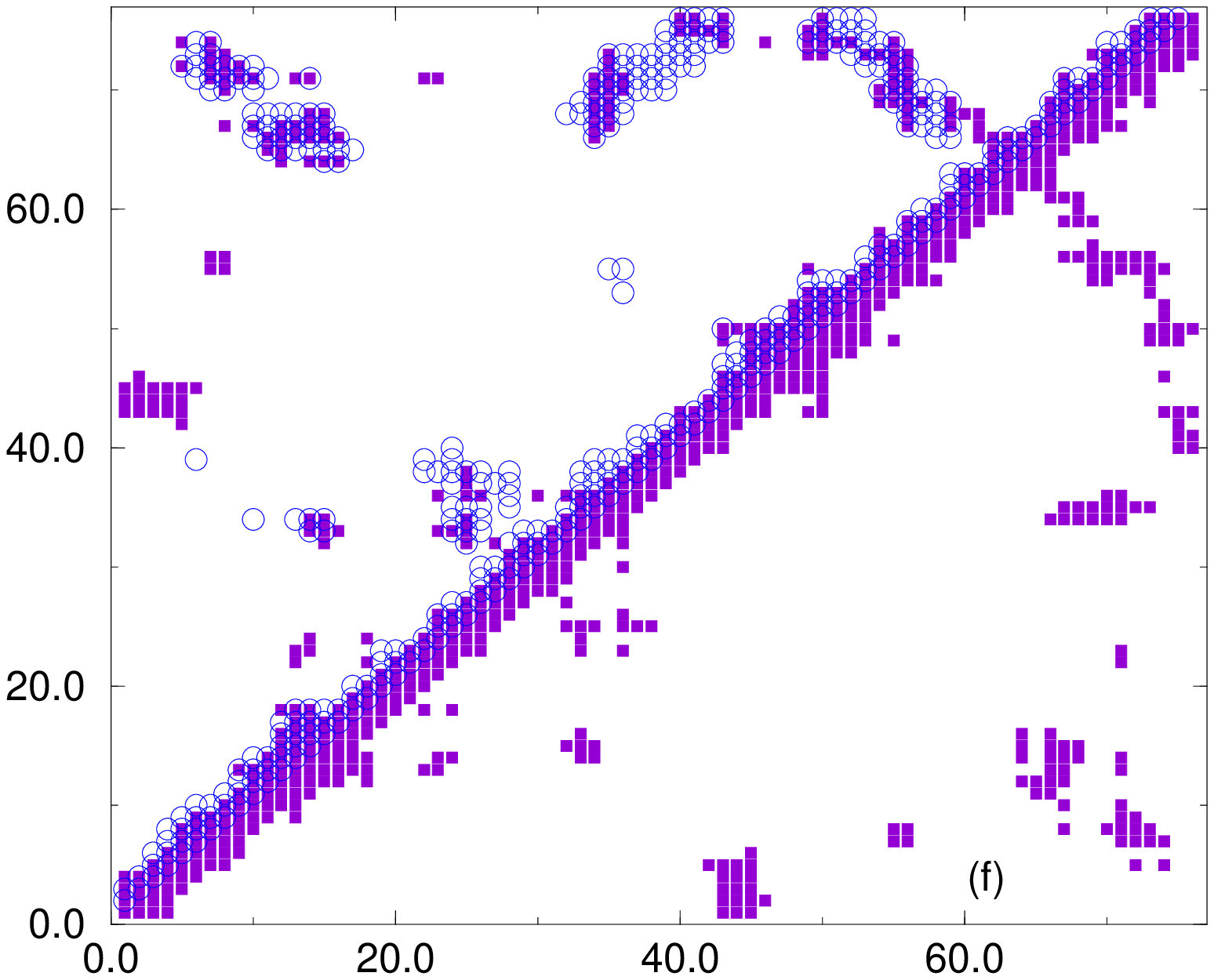,height=6.0cm,angle=0}}
\caption{
(a) Contact map for the native state of protein ubiquitin (1ubq).
There are 292 non-nearest neighbors contacts.
Region 1 is an $\alpha$ helix, region 2
a parallel $\beta$ sheet, and region 3 an antiparallel $\beta$ sheet.
(b) Contact map after a step of the non-local dynamics.
(c) After a step of the local growth dynamics. The untwisting of a helix
    is shown in the box.
(d) After a step of the local conservative dynamics.
(e) After reconstruction.
(f) After final minimization in real space.
In (c-f) squares are the current map and circles the previous one.
(Reprinted with permission from Vendruscolo and Domany (1998a).
Copyright 1998 Current Biology.) 
}
\label{fig:1ubq}
\end{figure}
\end{minipage}

\section{An approximation for the free energy}
\begin{quote}
{\em 
First, we introduce the exact free energy of a contact map and
discuss two simple approximations to it. Second, we present a method
to derive energy parameters based on perceptron learning.
}
\end{quote}

\subsection{The (free) energy associated with a contact map}
As explained above, many microscopic configurations of a protein with
sequence {\bol A} $=~~(a_1,a_2,a_3,...a_N)$ are
characterized by the same contact map {\bf S}. We now show that one can define
an {\it exact free energy}, ${\cal H} \SA$ , 
for the assignment of {\bf S} to the
sequence {\bf A}. Denote by $\C$ a micro-state of the system, specified by
the coordinates of all  atoms of the protein
(and of the solvent and any other relevant molecules). The true, microscopic
energy of this configuration is $E(\C)$. In thermal equilibrium each micro-state
appears with a probability proportional to the corresponding Boltzmann weight
$e^{-\frac{1}{kT}E({\C})}$.

The {\it free energy}
${{\cal H} \SA}$ (to which we refer simply energy)
associated with sequence {\bf A} and map {\bf S} is defined as follows:
\beq
{\rm Prob({\bol S})} \propto e^{- {\cal H} \SA} =
\sum_{\C} e^{-\frac{1}{kT}E({\C})} \Delta({\C},{\bol S})
\label{eq:free}
\eeq
where
\[
\Delta(\C,{\bol S})= \left\{ \begin{array}{ll}
                        1 & \mbox{if {\bol S} consistent with $\C$ } \\
                        0 & \mbox{otherwise}
\end{array}
\right.
\]
$\Delta(\C,{\bol S})$
is a ``projection operator'' that ensures that only 
those configurations ${\cal C}$ whose contact map is {\bol S} 
contribute to the sum (\ref{eq:free}). In other
words, only those micro-states whose contact map is {\bf S} contribute to the
sum and hence to  ${{\cal H} \SA}$.

This definition of the (free) energy of a map is exact;
it is nothing but the negative log of the {\it probability} 
of observing the map {\bf S} for sequence {\bf A}. 
Therefore ${\cal H} \SA$  has an important
property; inasmuch as the native fold's contact map, ${\bf S}_0$, has the
highest probability of appearing the corresponding (free) energy,
${\cal H} ({\bf S}_0, {\bf A})$ is the {\it 
lowest} among all possible ${\cal H} \SA$, i.e.
\beq
{\cal H} ({\bf A},{\bf S}_0) < 
{\cal H} ({\bf A},{\bf S}) \qquad \qquad \forall
{\bf S} \neq {\bf S}_0
\label{eq:lowest}
\eeq

The main problem with this exact energy is that the sum (\ref{eq:free})
is impossible to carry out. Therefore, one takes various 
phenomenologically motivated guesses for
the form of ${\cal H} \SA$, that presumably would have been obtained
had the sum  been carried out. This approach is related in spirit to the
phenomenological Landau-Ginzburg type free energy used in several areas of
condensed matter physics. We also start from the simplest approximate form
for this complicated function - that of the {\it pairwise contact energy}:

\beq
{{\cal H}}^{pair} \SA = \sum_{i<j}^N {{\bol S}_{ij}}  {w( a_i,a_j)}
\label{eq:pair}
\eeq
That is, if there is a contact between residues $i$ and $j$, the
parameter $w(a_i,a_j)$, which represents the energy gained
by bringing amino acids $a_i$ and $a_j$ in contact is added to the energy.

Another term the has been introduced (Mirny and Domany, 1996) 
in the same spirit is a {\it hydrophobic}, (or solvation) term, of the form 
\beq
{{\cal H}}^{hydro} \SA = \sum_{i}^N \left[\beta(a_i) \left(
\sum_{k \neq i}^N {\bol S}_{ik} - n(a_i) \right)^2  \right]
\label{eq:hydro}
\eeq
Here $n(a)$ is the desired number of contacts that amino acid $a$ should have -
as obtained from the PDB, for example. A hydrophobic residue will have a larger
value of $n$ than a hydrophilic one. The term penalizes the configuration 
for deviations of the actual number of contacts of $a_i$, as read from 
the map {\bf S}, from the desired $n(a_i)$. 
Whereas the terms in ${{\cal H}} ^{pair}$ 
are specific to the residue with which $a_i$ is in contact, here $a_i$ does
not ``care'' who are its partners. 
There may be some ``double counting'' in using
both energy terms, but this should be compensated in the way one determines the
parameters $w(a,b)$ and $\beta(a)$, the latter of which measure the strength of 
these hydrophobic terms. In our approach we determine the parameters 
by {\it learning}, with no recourse to a microscopic interpretation, so that 
we should not worry about double counting.    

\subsection{Optimization}

In the energy function of the form of Eq. (\ref{eq:pair})
one has to decide which set of parameters ${\bf w}$ to use.
The first idea was to derive a particular set of energy parameters
from amino acid pairing frequencies
observed in available crystallographic structures 
(Tanaka and Scheraga, 1976; Miyazawa and Jernigan, 1985).

An alternative method was proposed later by
Maiorov and Crippen (1992). Energy parameters are derived
by requiring that the observed native structures should be 
the lowest in energy among an ensemble of alternative conformations.
Given the native contact map ${\bf S}_0$, 
for each alternative map ${\bf S}$ they wrote an inequality
\begin{equation}
{\cal H}^{pair}({\bf A},{\bf S}_0,{\bf w}) < 
{\cal H}^{pair}({\bf A},{\bf S},{\bf w})  \;,
\label{eq:optimization}
\end{equation}
and obtained the set ${\bf w}$ from the solution of such inequalities.
It was also realized that parameter derivation can be formulated
as an optimization problem. 
For a fixed set of sequences with their native maps and for a fixed
set of alternative contact maps,
one defines a cost function in parameter
space and looks for the set of parameters of ``minimum cost''.
Goldstein {\em et al.} (1992) 
maximized the ratio $R$ between the width of
the distribution of the energy and the average energy difference
between the native state and the unfolded ones.
More recently Mirny and Shakhnovich (1996)
expressed the $Z$ score as a function of the energy parameters
which were then derived by optimization.

\subsection{Learning the energy parameters by a perceptron}

We set out to determine the energy parameters 
on the basis of the {\bf basic requirement} (\ref{eq:lowest}).
The basic requirement expresses the condition that
the energy should attain
its lowest value at the {\it true native map}.
It can be stated by posing the following question:
\begin{quote}
{\em Is it possible choose energy parameters such that 
for all the proteins in the database the native states
have the lowest energy among all
possible decoys?}
\end{quote}

We choose one protein (or more) with a known native map {\bf S}$_0$ 
and generate for
it a large set of decoys {\bf S}$_\mu$; then we look for 210 
contact parameters for which the inequalities (\ref{eq:optimization})
hold for all {\bf S}$_\mu$.

The basic attribute of ${{\cal H}}^{pair}$ that we use is its linearity in the 
contact energies $w(a,b)$; 
for any map ${\bf S}_\mu$ the energy (\ref{eq:pair})
is a linear function of the 210 contact energies that can appear and it
can be written as 
\begin{equation}
{\cal H}^{pair}({\bf A},{\bf S}_\mu, {\bf w}) =
\sum_{c=1}^{210} N_c({\bf S}_\mu) w_c
\label{eq:newener}
\end{equation}
Here the index $c=1,2,...210$  labels the different contacts that can appear
and $N_c({\bf S}_\mu)$ is the total number of contacts of type $c$ that actually
appear in map ${\bf S}_\mu$.
The difference between the energy of this map and the native one is therefore
\begin{equation}
\Delta {\cal H}^{pair}_\mu =
\sum_{c=1}^{210} w_c x^\mu_c = {\bf w} \cdot {\bf x}_\mu
\label{eq:Ediff}
\end{equation}
where we used the notation
\begin{equation}
x^\mu_c = N_c({\bf S}_\mu)- N_c({\bf S}_0)
\label{eq:Ndiff}
\end{equation}
and ${\bf S}_0$ is the native map.

Also the difference in the hydrophobic energy 
between the decoy ${\bf S}_\mu$ and the native map can be cast in a
linear form.  To see this, note that
\begin{equation}
{\cal H}^{hydro} ({\bf a}, {\bf S}_\mu, {\bf v}) =
\sum_{c=1}^{20} M_c({\bf S}_\mu) v_c \; .
\end{equation}
In this case the index $c$ runs over the 20 species of amino acids and
\begin{equation}
M_c({\bf S}_\mu)
=\sum_{i=1}^N \left[ 
\sum_{k \neq i}^N {\bol S}_{ik,\mu} - n(a_c) \right]^2 
\delta(a_i,c) \; ,
\end{equation}
where $\delta(a_i,c)=1$ if $a_i=c$ and 0 otherwise.

The difference between the hydrophobic
energy of this map and the native one is therefore
\begin{equation}
\Delta {\cal H}^{hydro}_\mu =
\sum_{c=1}^{20} v_c y^\mu_c = {\bf v} \cdot {\bf y}_\mu
\end{equation}
where we used the notation
\begin{equation}
y^\mu_c = M_c({\bf S}_\mu)- M_c({\bf S}_0)
\end{equation}

In the presence of the hydrophobic term,
the inequality (\ref{eq:lowest}) can then be written as
\begin{equation}
\Delta {\cal H}^{pair}_\mu + \Delta {\cal H}^{hydro}_\mu = 
{\bf w} \cdot {\bf x}_\mu + {\bf v} \cdot {\bf y}_\mu =
{\bf u} \cdot {\bf z}_\mu  > 0 \; ,
\label{eq:optz}
\end{equation}
where we introduced the following 230-components vectors
\begin{equation}
\begin{array}{ll}
{\bf z} = (x_1,\ldots,x_{210},y_1,\ldots,y_{20}) \\ 
{\bf u} = (w_1,\ldots,w_{210},v_1,\ldots,v_{20}) \; .
\end{array}
\end{equation}
The total energy is linear in the parameters 
${\bf w}$ and
${\bf v}$.

We denote the normalization factor of the vector ${\bf z}_\mu$ by
\begin{equation}
Z_\mu = \sum_{c=1}^{230} \left( z^\mu_c \right) ^2  \; ,
\label{eq:norm}
\end{equation}
and from here on we set
$Z^\mu_c \leftarrow z^\mu_c/Z_\mu^{1/2}$,
so that both ${\bf z}$ and ${\bf u}$
are normalized to 1.

Once the requirement (\ref{eq:optimization}), 
has been expressed using an energy in the form (\ref{eq:optz}), 
the question whether it does or does not have a solution reduces to deciding
whether a set of examples is learnable by a perceptron
(Rosenblatt, 1962).
Every candidate contact map provides a pattern for the training session. 

The vector ${\bf u}$ is ``learned'' in the course of  a training session.
The $P$ patterns are presented cyclically; after presentation of pattern $\mu$
the weights ${\bf u}$ are updated according to the following learning rule:
\begin{equation}
{\bf u}^{\prime} = \left\{ 
\begin{array}{ll}
({\bf u} + \eta {\bf z}_{\mu})/
                       |{\bf u} + \eta {\bf z}_{\mu}| \qquad & 
{\rm if} \qquad  {\bf u} \cdot {\bf z}_{\mu} <0 \\
 & \\
~~ {\bf u} & {\rm otherwise}
\end{array}
\right.
\label{eq:learnrule}
\end{equation}
This procedure is called learning since when the present $\bf u$ misses
the correct ``answer'' $h_\mu >0$ for example $\mu$, all weights are 
modified in a manner that reduces the error. No matter what
initial guess for the $\bf u$ one takes,
a convergence theorem guarantees that if a solution ${\bf u}$ exists,
it will be found in a finite number of training steps.
Since the parameters ${\bf v}$ have to be positive, we introduced the
following trick. We added
20 fictitious examples ${\bf z}$ which are vectors of zeros except
the component $y_i, i=1,\ldots,20$ which is set to 1.

Different choices are possible for the parameter $\eta$.
Here we use the learning rule introduced by 
Nabutovsky and Domany (1991)
since it allows, at least in principle, to assess whether the problem
is learnable.
The parameter $\eta$ is given at each learning step by
\begin{equation}
\eta = \frac { -h_{\mu} + 1/d } { 1-h_{\mu}/d }
\end{equation}
where the parameter $d$ (called despair) evolves during learning
according to
\begin{equation}
d^{\prime} = \frac { d + \eta } { \sqrt{1 + 2 \eta h_{\mu} + \eta ^2} } \; .
\end{equation}
Initially one sets $d=1$.

The training session can terminate with only two possible outcomes.
Either a solution is found (that is,
no pattern that violates condition (\ref{eq:optz}) is found in a cycle),
or unlearnability is detected. The problem is {\it unlearnable} if
the despair parameter $d$ exceeds a
critical threshold
\begin{equation}
d > d_c = \sqrt{M} \left[ 2 {Z}_{max} \right]^{M/2} \; ,
\label{eq:dc}
\end{equation}
where $M$ is the number of components of ${\bf w}$,
and ${Z}_{max}$ is the maximal value of the normalization factors
(see Eq (\ref{eq:norm})). This value of $d_c$ is easily derived
for examples of the type of Eq (\ref{eq:Ndiff}), using the same method
as given by Nabutovsky and Domany (1991).

It is easy to see that ${\bf u}^*=({\bf w}^*,{\bf v}^*)$ is a solution
of the system (\ref{eq:optz})
if and only if  ${\bf u}^*_\lambda=({\bf w}^*,\lambda {\bf v}^*)$ is
a solution of
\begin{equation}
{\bf w} \cdot {\bf x}_\mu + 
\frac{1}{\lambda}\; {\bf v} \cdot {\bf y}_\mu > 0 \; .
\label{eq:lambda}
\end{equation}
The outcome of the learning process {\bf does not} depend on the
initial choice of $\lambda$. However,
the learning time usually depends on $\lambda$ which has to be chosen
conveniently. A useful condition to set $\lambda$ is 
to obtain on average $|{\bf x}| \sim | {\bf y}/\lambda | $.

\section{Results}
\begin{quote}
{\em 
We prove in an extensive number of situations
that the pairwise contact approximation both when
alone and when supplemented with a hydrophobic term is unsuitable
for stabilizing proteins' native states.
}
\end{quote}

Whether or not the basic requirement (\ref{eq:lowest})
can be satisfied depends on several factors. 
We will discuss the dependence upon the
following issues:
\begin{enumerate}
\item
The definition of contact.
\item
The assignment of the contact length $R_c$.
\item
The number $M_p$ of proteins in the database.
\item
The method used to produce decoys.
\end{enumerate}

\subsection{Threading}

We first discuss the dependence of learnability on $M_p$ and $R_c$
using the all atoms definition of contact and producing decoys
by gapless threading.
Once the contact maps of the $M_p$ proteins in the database are obtained,
decoys are generated for a given sequence of length $N$ from
the structures of  proteins of lengths $N^\prime$ ($>N$) by selecting
submaps of size $N\times N$ along the main diagonal of the
contact map of the longer protein.
For each definition of contact that we considered we found two ``phases''.
There is a region in the $(R_c,M_p)$ in which
the problem is {\em learnable}; that is, there is a set ${\bf w}$
of pairwise contact energy parameters that stabilize simultaneously
all the native maps in the set. On the other hand, outside this region
(e.g. for fixed $R_c$ and large enough $M_p$)
the problem is ${\em unlearnable}$, and no set ${\bf w}$
exists. 

Without doing any calculation, we can predict a few general features
of the $(R_c,M_p)$ phase diagram.
Having set a definition of contact (e.g. the all atoms one)
one can plot the distribution of distances between amino acid pairs.
Choosing $R_c$ smaller than the smallest observed distance
would result in contact maps with no contacts, independently of
the conformation. No set of energy parameters can then discriminate
the native map from the decoys.
Similarly, choosing $R_c$ larger than the largest observed distance
would result in contact maps with all the entries set to 1.
In this case again no discrimination is possible.
Thus we expect to find a window of learnability in $R_c$.
It is also reasonable to expect that such a 
window will shrink with increasing $M_p$.
The problem is thus reduced to investigate whether such window remains open
for an arbitrary large value of $M_p$, or it closes for $M_p$ large enough.

The boundaries of the region of learnability must be interpreted
in a probabilistic sense.
At given $M_p$ and $R_c$, learnability depends on the particular choice
of the proteins in the database.
In principle one should define $P(R_c,M_p)$, the probability
for a randomly extracted set to be learnable at $(R_c,M_p)$.
The boundary is then defined by $P(R_c,M_p)=const$.
In the present study,
we chose not to give a precise numerical estimation of $P(R_c,M_p)$, 
which would require many extractions of sets of $M_p$ proteins from the PDB
and would be numerically intensive.
We are interested in establishing the {\em existence} of the
boundary rather than in its precise determination. Hence we will show that
it is possible to find unlearnable datasets when their size is large enough.

\begin{figure}
\centerline{\psfig{figure=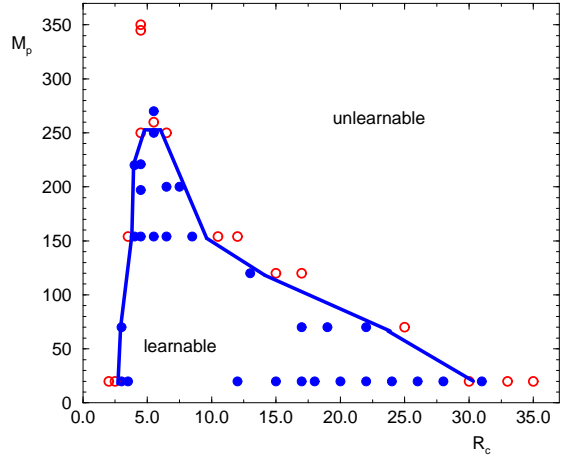,height=7.0cm,angle=0}}
\caption{
Region of learnability for the all atoms definition of contact.	Several
sets of $M_p$ proteins were generated for each value of $R_c$.
Full circles indicate that all the sets considered for a particular
value of $(R_c,M_p)$ were learnable; otherwise we use open circles.
}
\label{fig:aa.phase}
\end{figure}

The approximate boundaries of the region of learnability
are shown in Fig. \ref{fig:aa.phase}.
The window of learnability shrinks to zero for $M_p$ above 270.
The precise value of the limit of learnability 
depends on the particular choice of proteins, and the 
fluctuations in $M_p$ are of the order of few tens of proteins.
Such fluctuations are larger than expected; they are due to a 
few proteins that are markedly more difficult to learn
than others, and their inclusion in the dataset lowers the chances
of learnability. For example 1vdfA and 1gotG are
are two such ``hard to learn'' 
proteins. 1vdfA
is a chain of 46 amino acids 
which forms a single $\alpha$ helix; 1gotG is a chain formed
by 2 $\alpha$ helices hinging at 90 degrees.

Each point shown in Fig. \ref{fig:aa.phase} is derived from a few (1 -3) 
randomly generated sets of $M_p$ proteins in each.
A full circle indicates that all the sets considered were learnable; open circles
signal that at least one set was unlearnable.
The largest fluctuations are found for small $M_p$ at the right boundary.
For example, for different sets with $M_p$=20
we found that the maximal $R_c$ of learnability
varies between 28 \AA \hspace{2pt} and 31 \AA.

In this study, we selected five sets of protein structures
from the PDB.
The sets ranged from 154 to 945 proteins.
The PDB is an archive of experimentally derived structures of proteins.
To date, 8295 coordinate entries are deposited.
It is known that the information contained in the
entire PDB is redundant and some is incorrect.
Routine methods are available to select subsets of non-homologous
proteins whose experimental structures
are determined reliably 
(Heringa {\em et al.}, 1992; Hobohm and Sander, 1994; Hooft {\em et al.} 1996).

Correlations between solutions at different $M_p$ and $R_c$ ranged
from 0.22 (for a solution at $R_c=7.5$ \AA \hspace{2pt} and $M_p=200$
with a solution at $R_c=4.5$ \AA \hspace{2pt} and $M_p=154$)
to 0.94, which is the typical correlation between two solutions of maximal
stability at $R_c=4.5$ \AA \hspace{2pt} and $M_p > 150 $.

These findings indicate that
with the all atoms definition of contact and 
the physically motivated choice of $R_c=4.5$ \AA \hspace{3pt} 
(Mirny and Domany, 1996)
it is possible to  stabilize simultaneously, with high probability,
about 200 randomly selected non homologous proteins against decoys
generated by gapless threading. 

\begin{figure}
\centerline{\psfig{figure=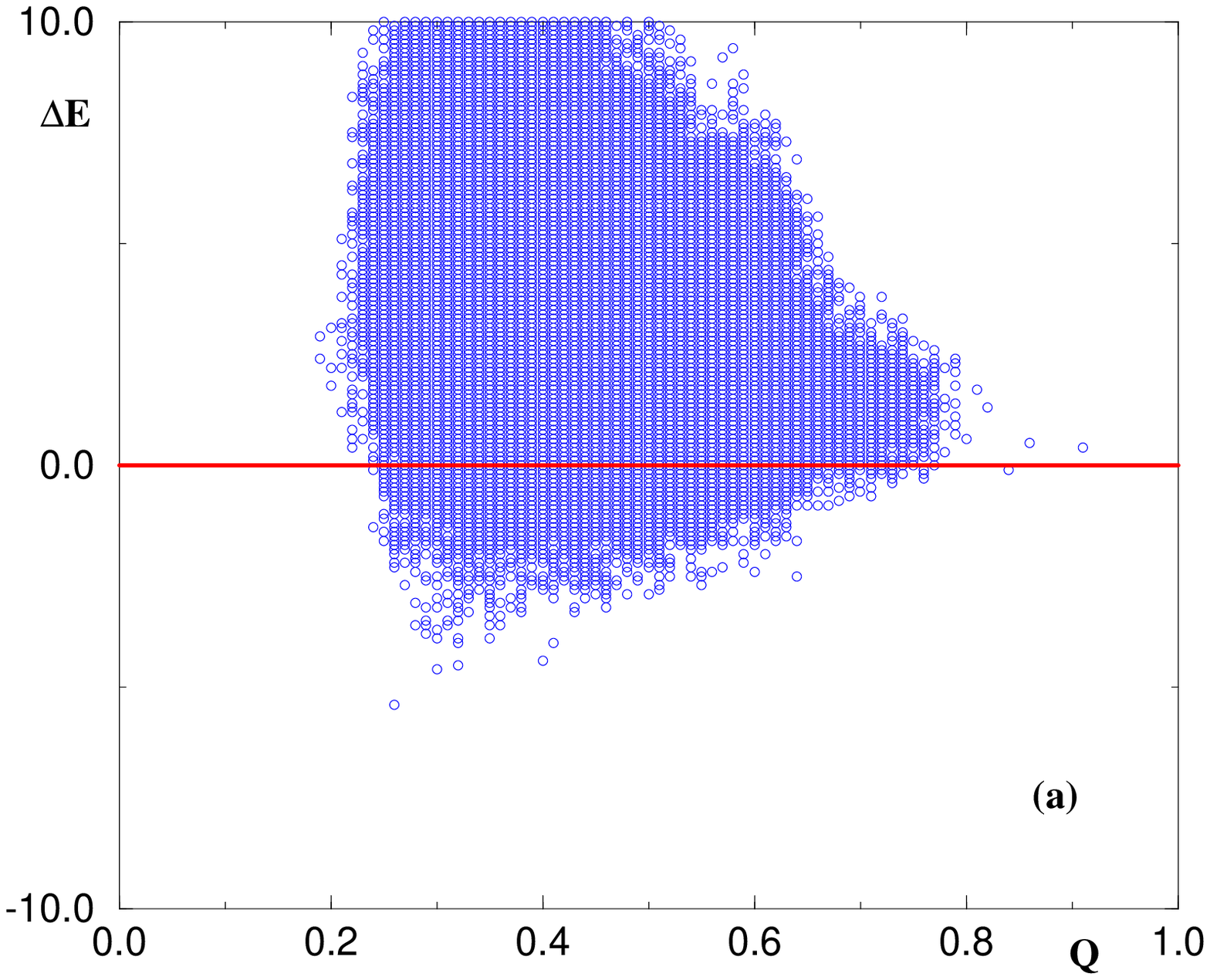,height=7.0cm,angle=0}
            \psfig{figure=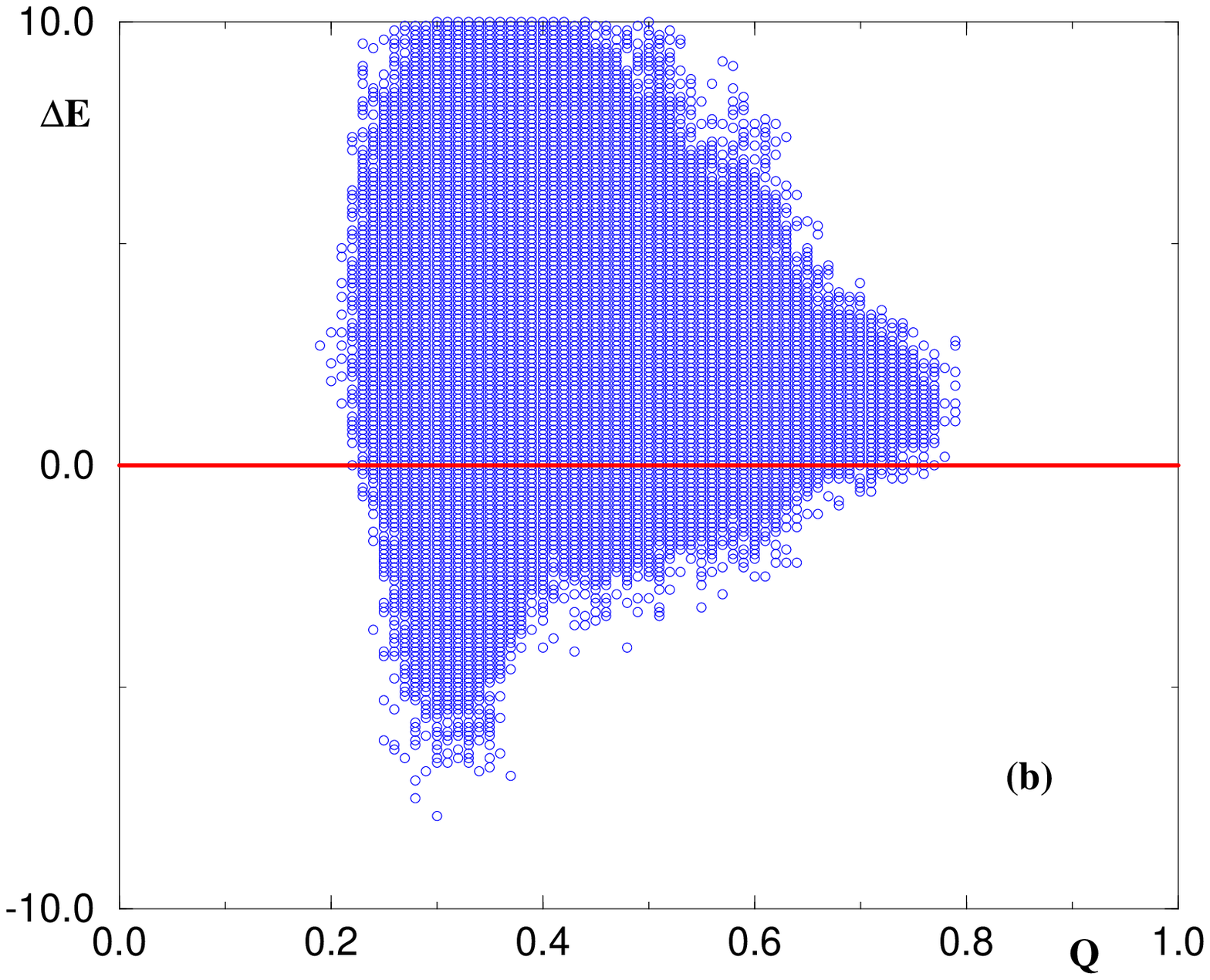,height=7.0cm,angle=0}}
\centerline{\psfig{figure=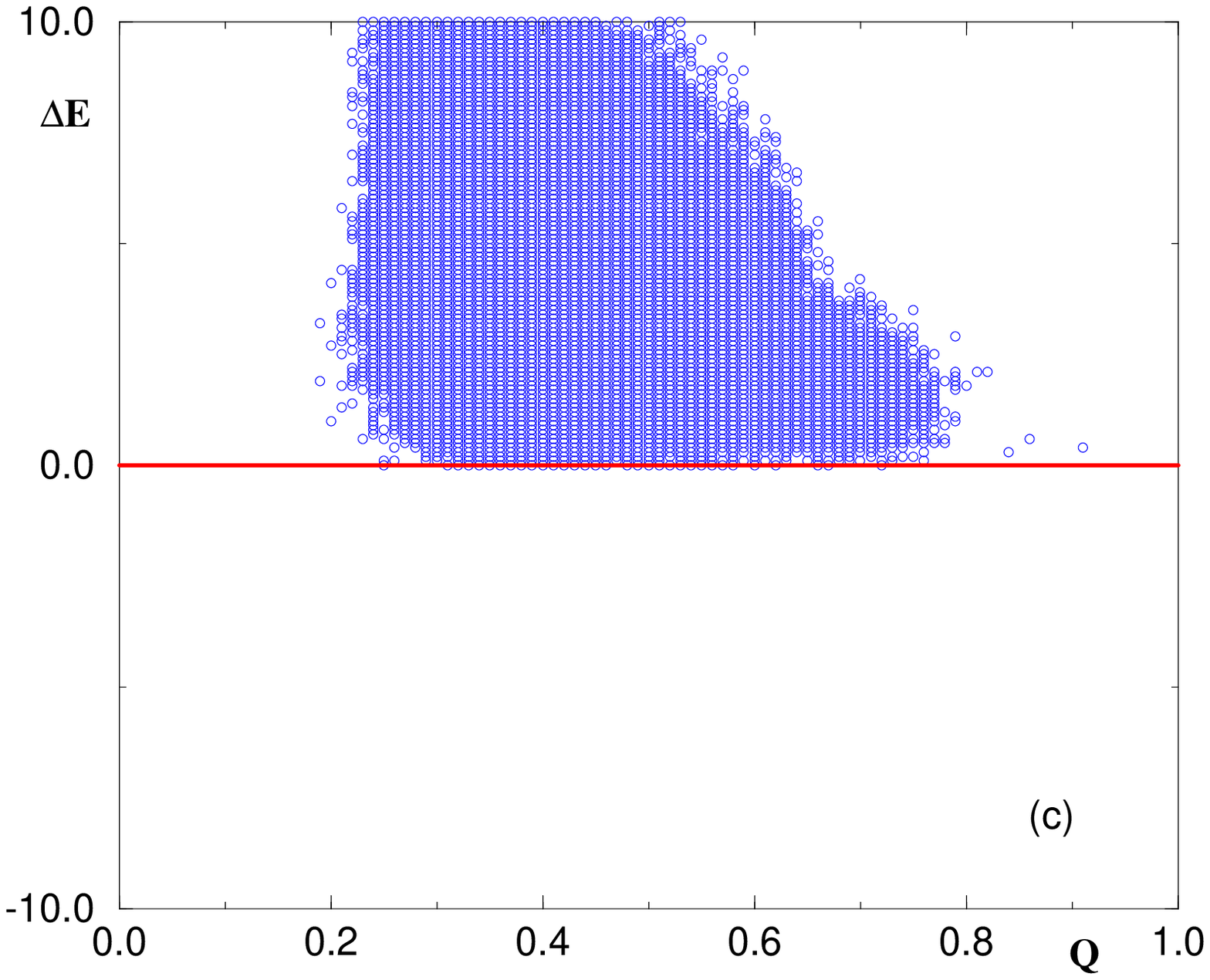,height=7.0cm,angle=0}
            \psfig{figure=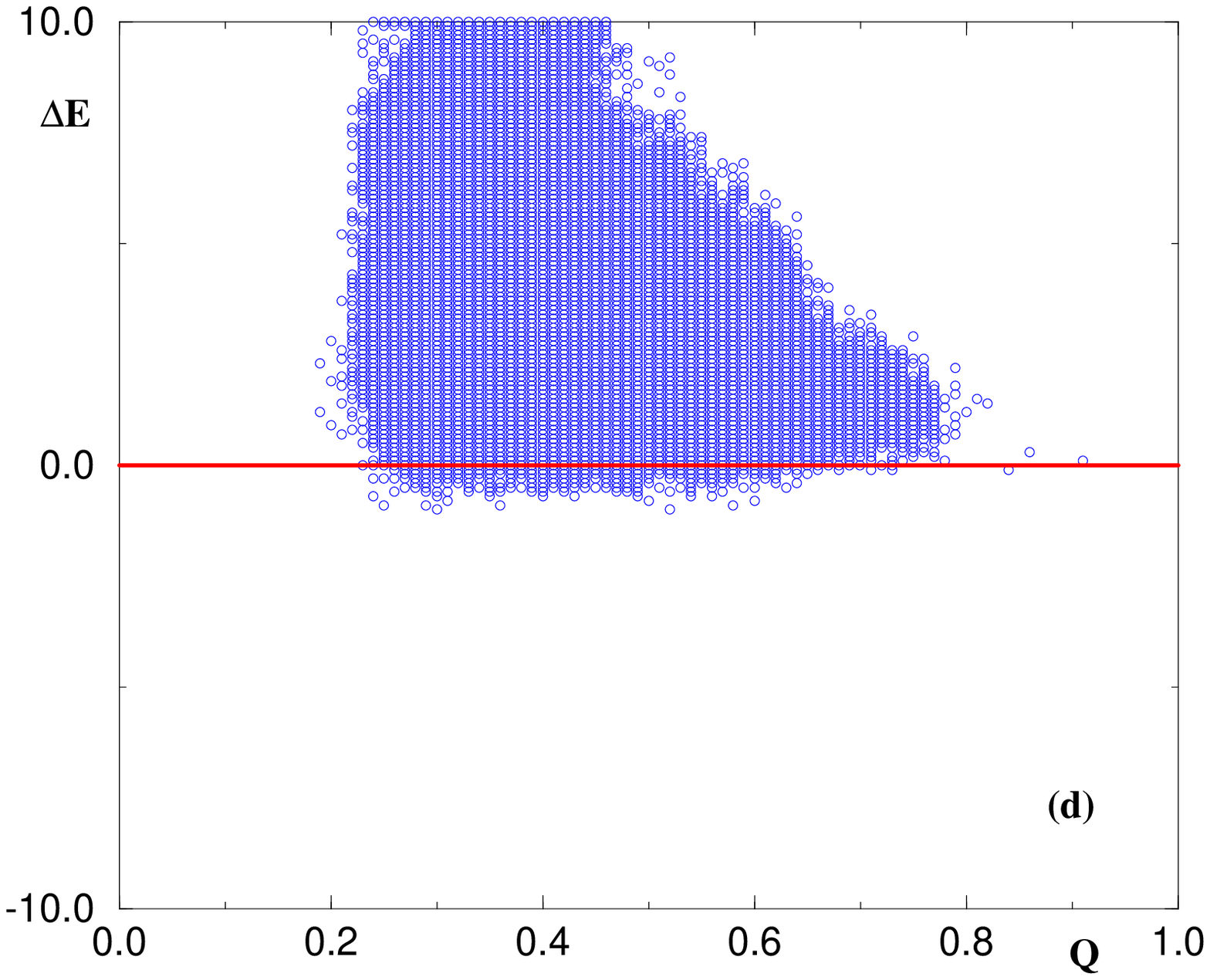,height=7.0cm,angle=0}}
\caption{
Scatter plot of the energy difference ${\cal H}$ between decoys and native state
and the overlap $Q$ with the native state,
(a) Learnable set of 123 proteins 
for which the energy ${\cal H}^{pair}$ was calculated
using the solution of an independent set of 197 proteins;
(b) Unlearnable set of 141 proteins
with the same initial set of energy parameters;
(c) The set of 123 proteins with the energy parameters obtained from learning;
(d) The set of 141 proteins with the energy parameters 
arrived at when unlearnability has been established.
}
\label{fig:eq}
\end{figure}

Which decoys are more challenging?
It is instructive to consider the overlap $Q$ between 
the contact map of the native state and of the decoys, 
which is defined as
\begin{equation}
Q = \frac{1} {\max(N_c^0,N_c^\mu)}
\sum_{h>k+1}^{N} {\bf S}_{hk}^0 {\bf S}_{hk}^\mu \; ,
\label{eq:overlap}
\end{equation}
where $N$ is the length of protein and $N_c^0$ and $N_c^\mu$
are the number of contacts in ${\bf S}^0$ and ${\bf S}^\mu$, respectively.
We considered, for the $C_\alpha$ definition and $R_c$=8.5 \AA,
two sets of proteins. The first, of 123 proteins, is learnable
and the second, of 141 proteins, unlearnable.
First, for each decoy we calculated, 
using as initial weights the solution of an independent set
of 197 proteins, the energy difference $\Delta {\cal H}$
with the native state and the overlap $Q$. 
In Figs. \ref{fig:eq}(a) and (b) we present scatter plots of
of $\Delta {\cal H}$ on $Q$ for our decoys.
The first question we asked is whether decoys of high overlap are the
more challenging ones.
To answer this question we repeated the learning procedure
by considering only decoys with $Q < Q_t$, where $Q_t$
is a threshold value for the overlap.
The set of 141 proteins was still unlearnable for $Q_t=0.6$.
It seems that the ``difficult'' decoys 
are spread over the entire range of $Q$.
Including all the decoys in the learning procedure we were able
to learn the set of 123 proteins, but not the set of 141 proteins.
As shown in Fig. \ref{fig:eq}(c), after learning the set of 123 proteins,
decoys of low energy are present in the approximate range $0.2<Q<0.8$. 
The important finding is that the unlearnable case
is not qualitatively different (see Fig. \ref{fig:eq}(d)).
Also in this case 
decoys of low energy are present in the range $0.2<Q<0.8$,
although now some of them have $\Delta {\cal H} < 0$.
The difference is that with the set of 123 proteins there are
805,938 decoys, whereas for the set of 141 proteins, 1,071,753.
With 210 energy parameters it is possible to satisfy the smaller 
set of inequalities but not the larger.
Decoys of arbitrary $Q$ enter in the learning process 
and one can argue that 
the lack of correlation between $\Delta {\cal H}$ and $Q$ is a major cause that
renders the problem unlearnable for large enough $M_p$
and, further, that an improved energy function
should first of all provide such a correlation.

To investigate the dependence of learnability on the definition of contact
we repeated the same analysis for two other definitions of contacts.
The first based on $C_\alpha$ and the second on all carbon atoms --
a contact was assumed if any two $C$ atoms of the two amino acids
were closer than $R_c$.
In both cases we found regions of learnability qualitatively similar to the 
one shown in Fig. \ref{fig:aa.phase}

To summarize, we showed that, no matter which definition of contact
and which $R_c$ are chosen, there is always a maximal number
(of the order of few hundreds)
of proteins that can be stabilized together.
Is this number large or small?
A definite answer is outside the limits of this study.
It is {\it small}
when compared with the total number of proteins existing in nature
(hundreds of thousands),
but it is fairly large when compared with the number of representative
proteins in the PDB (also of the order
of few hundreds (Hobohm and Sander, 1994).
We choose to generate decoys by gapless threading
since it is a very efficient way to obtain decoys.
To further generalize our conclusion,
in the following sections we consider a more refined way to obtain decoys,
based on energy minimization in the space of contact maps.

\subsection{Crambin}

\begin{figure}
\centerline{\psfig{figure=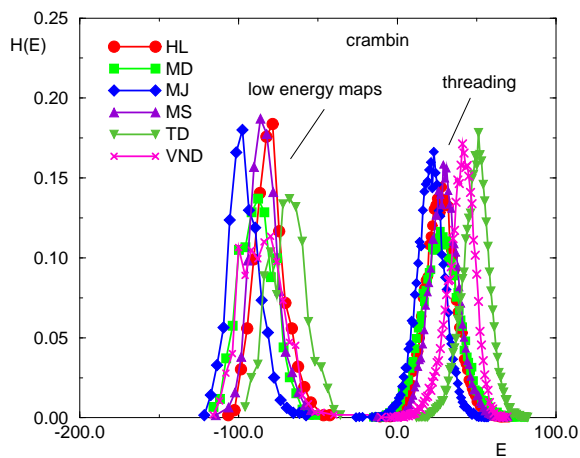,height=7.0cm,angle=0}}
\caption{
Histograms that demonstrate the difference in energy between
ensembles of contact maps obtained by threading and by energy minimization,
shown for different contact energy parameters:
(VND) as obtained in this work, by finding a solution for
a threading ensemble;
(HL): (Hinds and Levitt, 1994);
(MD): (Mirny and Domany, 1996);
(MJ): (Miyazawa and Jernigan, 1996);
(MS): (Mirny and Shakhnovich, 1996);
(TD): (Thomas and Dill, 1996).
The energy parameter sets were shifted and rescaled to obtain
$\langle {\bf w} \rangle =0$ and
$\langle {\bf w}^2 \rangle - \langle {\bf w} \rangle ^2 =1$
(averages are over the 210 energy parameters).
Energies were shifted
to set the  native state to $E$=0.}
\label{fig:histogram}
\end{figure}

The conclusion drawn from threading is that there is a maximal number
of proteins that can be stabilized together using the pairwise
contact approximation to the energy.
One can ask a more limited question, namely if it is possible to fine tune
energy parameters to stabilize just one protein, or possibly a set of proteins
belonging to the same structural family.
Threading offers poor contenders to the native state.
Better candidates for the ground state are produced
by contact map dynamics, the method presented in Sec. \ref{sec:dynamics},
which explores in an efficient way the space of contact maps.
For a particular protein -- crambin --
we show in Fig. \ref{fig:histogram} that the decoys produced by threading
are far less a challenge than those produced by contact map dynamics.
The set of decoys obtained by threading can be learned --
it is possible to produce a set of pairwise contact energy parameters
that stabilizes crambin in a threading experiment.
Here we ask the same question for decoys obtained by contact map dynamics
(see point 4. at the beginning of this Section).

Crambin (Teeter {\em et al.}, 1993)
is a protein of length $N=46$. We constructed
its native map by taking the coordinates of the 
$C_{\alpha}$ atoms from the PDB and using a threshold
of 8.5 \AA \hspace{3pt} to define contacts.
In the crambin chain, 
5 out of the 20 amino acids do not appear and 3 appear only once. 
Thus, among the corresponding 210 possible contact energies,
only 117 parameters can effectively enter the energy (\ref{eq:pair})
for any set of candidate maps.
These parameters form a 117-component vector $\bf w$.
The native map contains 187 non-nearest neighbor contacts, 
which involve only 72 of the 117 possible contact energy parameters.
We summarize here our main result about the question we have addressed 
in the present work. We will present below decisive 
evidence supporting our main conclusion: 
\begin{quote}
the problem of fine tuning the pairwise contact energy parameters to stabilize 
the native state of crambin is {\em unsolvable}.
\end{quote}

\subsubsection{Learning the pairwise contact term}

In an unlearnable case there exist  sets of examples for which no solution
can be found;
for large enough $P$  the training set will include, with non-vanishing
probability, such an unlearnable subset.

The condition that has to be met
(Nabutovsky and Domany, 1991) to establish unlearnability, 
is that the despair $d$, recorded during the learning process,
should exceed the critical value $d_c$. 
According to Eq (\ref{eq:dc}), for $M=117$, 
we typically get a critical despair $d_c \simeq 10^{163}$.

In perceptron learning of $P>2M$ randomly generated examples 
(which is an unlearnable problem
(Cover, 1965; Gardner, 1988) for large $M$), 
$d$ grows {\em exponentially} with learning time 
(Nabutovsky and Domany, 1991).
Here, due to the specific non-random nature of the learning task,
we need much more than $P=2M=234$ examples.
Learning is realized in an iterative manner.
Starting with an initial set of energy parameters we
generate a set of examples that are then learned; the new set of energy
parameters is used to generate new examples and so on.
We will refer to each such iteration as an {\em epoch}.
It was necessary to generate $P$=298710 examples in 11 epochs.
We underscore the fact that each example was obtained by an energy minimization
procedure, using the method discussed in Sec. \ref{sec:dynamics}.
Learning this set of examples is an unsolvable task, 
proved by the divergence of the despair $d$.
We reached $d>d_c$ after approximatively 37500 learning cycles.
To speed up the procedure, without affecting the final result,
we selected the 10000 examples at epoch 11 that had the lowest energy
according to the energy parameters obtained from learning at epoch 10.

\subsubsection{Energies of the false ground states}
\label{sec:crambin.pair.fold}.

\begin{figure}
\centerline{\psfig{figure=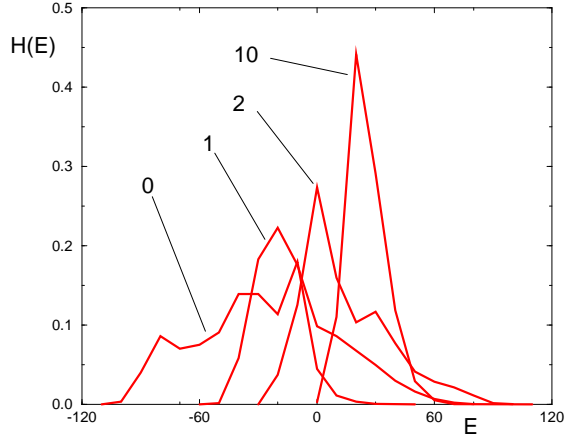,height=7.0cm,angle=270}}
\caption{
Normalized histogram $H({\cal H})$ of the energies of the contact maps
at epochs $t=0,1,2$ and 10. The energy scale is shifted so that
the native contact map energy is 0.
(Adapted with permission from Vendruscolo and Domany (1998b). 
Copyright 1998 American Institute of Physics.)
}
\label{fig:he}
\end{figure}

\begin{figure}
\centerline{\psfig{figure=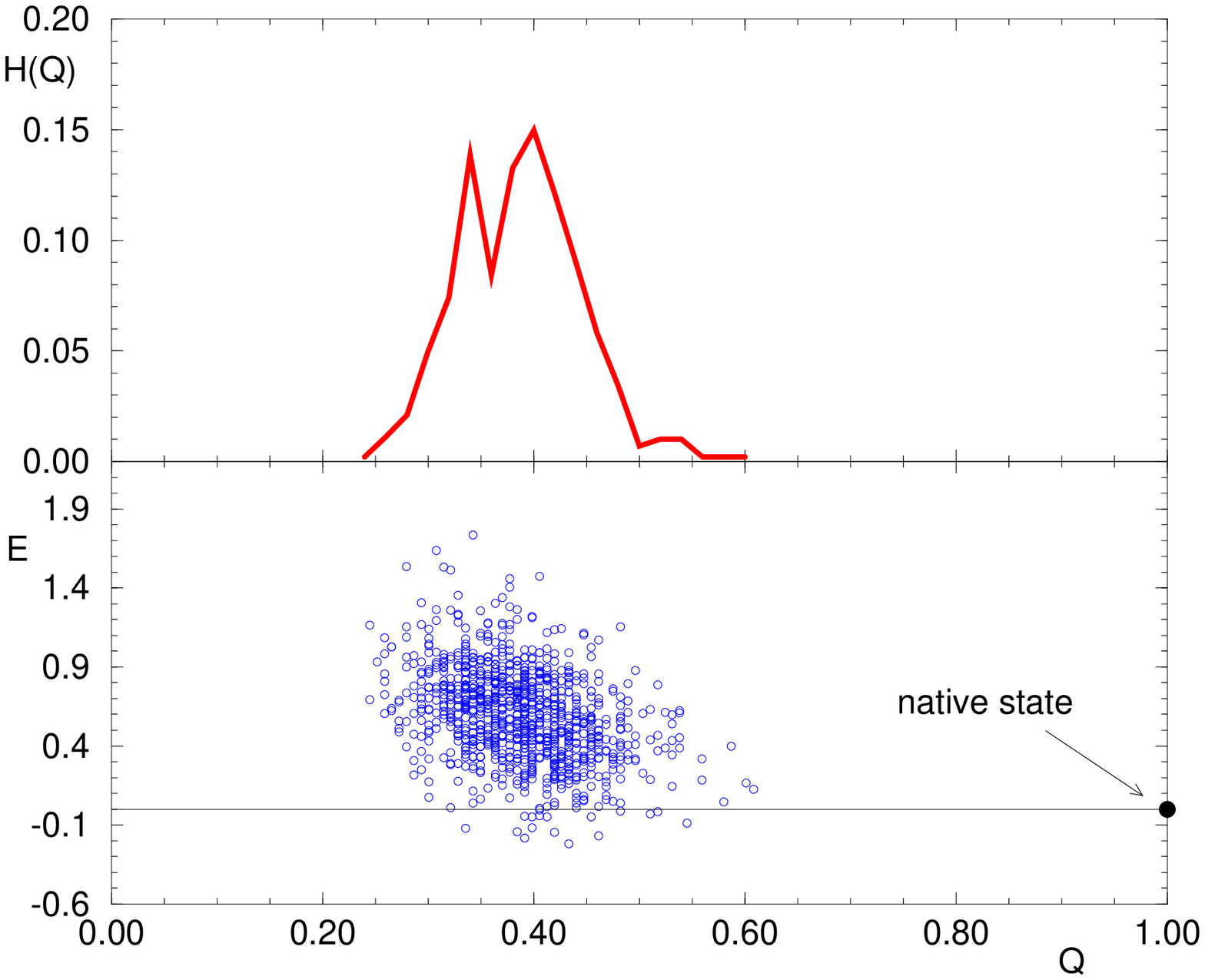,height=7.0cm,angle=0}
            \psfig{figure=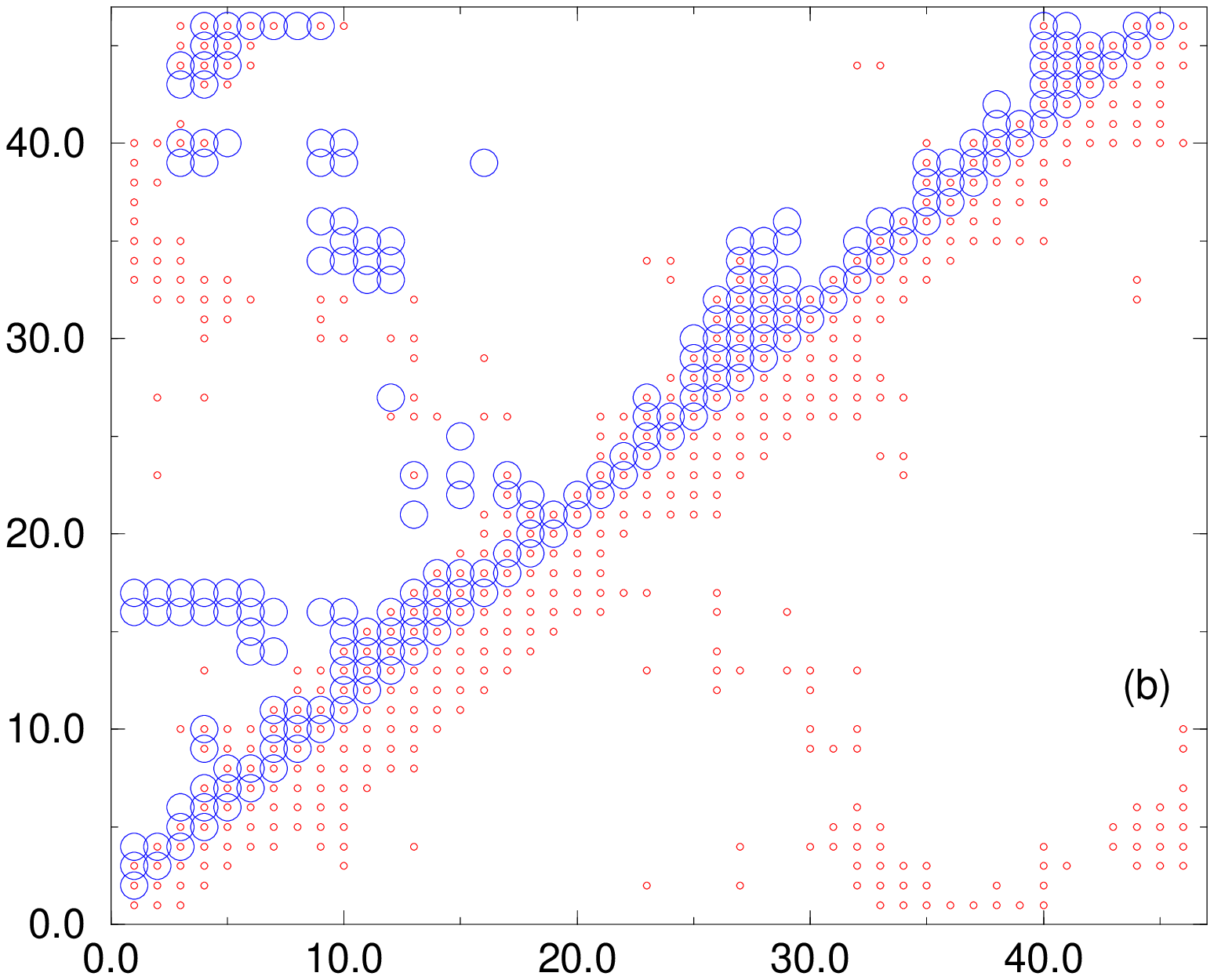,height=7.0cm,angle=0}}
\caption{
(a)
 Result of the folding experiment
 on crambin described in the text in which we generated 1000 low energy decoys.
 $Q$ is the fraction of correctly
 predicted contacts and ${\cal H}$ is the energy difference between the decoy
 and the native state.
(Adapted with permission from Vendruscolo and Domany (1998b). 
Copyright 1998 American Institute of Physics.)
(b)
 Typical low energy map obtained during the simulation. Small dots
 indicate contacts in the native map, open circles contacts in the
 simulated map.
}
\label{fig:qe}
\end{figure}

Even though the problem is unlearnable, our procedure
produces contact energies that have several appealing features.
The first is that whereas for the existing contact
potentials it is very easy to find maps whose energy is below that of
the native map, with the $\bf w$ obtained after several training epochs
this becomes a difficult (albeit possible) task.
With the initial energy parameters the vast majority of the contact maps
that are generated have a lower energy than the native contact map.
As can be seen from Fig. \ref{fig:he}, where the energy scale is shifted
so that the native contact map has always zero energy,
for increasing epoch index, the energy distribution shifts to the right
and becomes narrower. Learning is thus accompanied by an improvement
of the $Z$-score, which is a commonly used estimator of the quality of a set
of energy parameters (Mirny and Shakhnovich, 1996).
Hence our learning procedure flattens the energy landscape,
reducing both the number of violating examples and their energy difference
with the true native state.
Another relevant question concerns the ``location'' of these false minima,
i.e. how different are the corresponding structures from the native one?
To study this, we reconstructed the three dimensional
conformations corresponding to the violating examples and
measured their average distance $D$ (see Eq. \ref{eq:distance})
from the native conformation.
We found that D does not decrease with the epoch index;
moreover false minima are
found at an approximate average D of 8 \AA \hspace{3pt}
at {\em any} epoch.
Only their number decreased significantly.
Two compact uncorrelated conformations are typically found
at a distance of 15 \AA, which is also the result we obtained
using other parameter sets taken from the literature.
Here, we define the overlap $Q$ between contact maps 
in a slightly different way as than in Eq. (\ref{eq:overlap})
\begin{equation}
Q= \frac{N_p}{N_c}
\label{eq:over}
\end{equation}
where $N_p$ is the number of contacts present
in the native map {\em and} in the predicted one
and $N_c$ is the total number of contacts of the native map
(contacts $(i,i+1)$ and $(i,i+2)$ are not counted).
We generated 1000 low energy contact maps, using
the set of energy parameters obtained at epoch 10.
The histogram of the fraction of contact maps with a given overlap $Q$
with the native state is shown in Fig. 5.
The distribution is peaked at around 0.40, which means that typically
we are able to correctly recover 40 \% of the native contacts.
We analyzed the distance D of the conformations corresponding to the maps
of Fig. \ref{fig:qe}. We found that in the worst cases conformations are found
at an approximate average D of 8 \AA \hspace{3pt}.
It is not possible to improve this result, since the contact
energy parameters cannot be optimized further.

Consistently with our conclusion that the pairwise contact energy approximation
is unsuitable for protein folding, Hao and Scheraga (1996)
showed that a more accurate parametrization of the energy
is capable of improving the performance in a folding simulation.
They found that ``when the crambin model is simulated to fold to the
low-energy state with the optimized energy parameters, the folded structure
is generally close to the target native structure'', 
although ``the energy of the target structure was never able to
separate completely from the continuous density of the nonnative states
through the optimization''.
They do find nonnative conformations with energy lower than the native state,
but the contact maps of these have, 
on the average, high overlap with the native map.

One should not be misled by similar work on {\em model} proteins 
where, using a pairwise contact energy function,
it is possible to discriminate the native state, 
either by optimization 
(Hao and Scheraga, 1996; Mirny and Shakhnovich, 1996; Hao and Scheraga, 1998)
or by imposing conditions analogous
to Eq (\ref{eq:optimization}) 
(Van Mourik {\em et al.}, 1998).
In this case,
a database of foldable sequences is designed using a pairwise contact potential
and subsequently a set of contact energy parameters is recovered.
Success in this case is possible because the contact energy of
Eq (\ref{eq:pair}) is the {\em exact} form of the free energy of the model.

\subsubsection{Learning the hydrophobic term}
We proved that the pairwise contact approximation is unsuitable
to stabilize the native contact map of crambin against
a set of decoys obtained by contact map dynamics. 
The next step we take is to improve it
by adding the hydrophobic term (\ref{eq:hydro}) and ask 
again if the native state of crambin can be assigned with the lowest energy.

We found that the previously used set of 298710 decoys
was learnable with the hydrophobic term. We increased the
number of decoys to 390117 in 15 epochs 
and in this way we obtained an unlearnable set.
After some trial we choose $\lambda=15$ in eq. \ref{eq:lambda}
since we found that the despair $d$ grows faster for this choice.

Introducing the hydrophobic term does not make the problem learnable.
However, it improves the situation depicted in Fig. \ref{fig:qe}.
We repeated the folding experiment previously done 
(see Sec. \ref{sec:crambin.pair.fold}).
We found that the typical value of $Q$ is larger,
as shown in Fig. \ref{fig:ig}a. The average value of $Q$ moves from 0.4
(pairwise term only) to 0.5 (hydrophobic term).
These two values should be compared with 0.2, which is the value
obtained for decoys produced by threading crambin into a set
of 456 proteins (the distribution is marked by a $T$ in the figure). 

\subsection{Immunoglobulins}

Crambin could be a peculiar case since it is a short protein, it has
only 15 species of amino acids, etc. To what extent the results
about unlearnability extends to other proteins ?
To answer to this question we undertook a study of a set of proteins
of the family of immunoglobulins.

An immunoglobulin molecule is formed by two light chains and two
heavy ones (Branden and Tooze, 1991)
held together by disulfide bridges
and by the packing of their respective domains. 
The light chain folds into two domains,
the variable $V_L$ and the constant $C_L$ and the heavy chain
folds into four domain, one variable $V_H$ and three constant 
$C_{H1}$, $C_{H2}$ and $C_{H3}$.
Sequence similarity in constant domains is typically above 35 \% 
and around 10 to 20 \% in variable domains.

We focused our study on the variable domains of the light chain $V_L$
which is formed by two $\beta$-sheets, one of four strands and the other
of six strands. The reason to choose immunoglobulins is mainly due
to the extremely rich amount of information available.
The sequence database of Kabat {\em et al.} (1991) now contains 
about 5700 different variable domains and the structure of about 140
are available form the PDB.

We asked the question if it is possible to stabilize
6 $V_L$ domains simultaneously. The domains chosen were
8fab A, 1baf L, 1cbv L, 1dba L, 2f19 L and 1fdl L,
whose respective lengths were 104, 107, 112, 107, 107 and 107 amino acids.

\subsubsection{Learning the pairwise contact term}
As opposed to the case of crambin, where only 117 pairwise contact species
are present, all the 210 possible ones are present for immunoglobulins.
To prove unlearnability, we adopted the same iterative procedure
used for crambin.  We generated 97309 examples, in 6 epochs.
To speed up the procedure we considered only a subset of decoys,
selected according to the following procedure.
Consider the component $c$ of the vector ${\bf x}^\mu$ 
of a particular decoy $\mu$.
If the numbers $N_c(S_\mu)$ and $N_c(S_0)$ are equal 
then $x_c^\mu=0$ and this particular decoy can never be used to update $w_c$.
If we initialize $w_c=0$ and use only decoys with $x_c=0$
then $w_c=0$ during the entire learning procedure.
We excluded all the decoys having contacts of 13 species,
in this way we remained with 35677 decoys with 197 contact species.
Obviously, for such a subset, adding the 13 energy parameters cannot
change the result, since they never enter in the calculation of the energy.
We further selected the 5000 decoys of lowest energy
(according to the energy parameters learned at epoch 5)
and performed the learning on this subset.
In this way the critical value of the despair was surpassed in an amenable
computer time.

\subsubsection{Learning the hydrophobic term}
The set of 97309 decoys was learnable by adding the hydrophobic term.
We enlarged it to 110576.
Using the same trick explained above, we excluded from the learning
13 pairwise contact energy terms.
This selection procedure identified a subset of 42798 decoys
for which we proved unlearnability.

\begin{figure}
\centerline{\psfig{figure=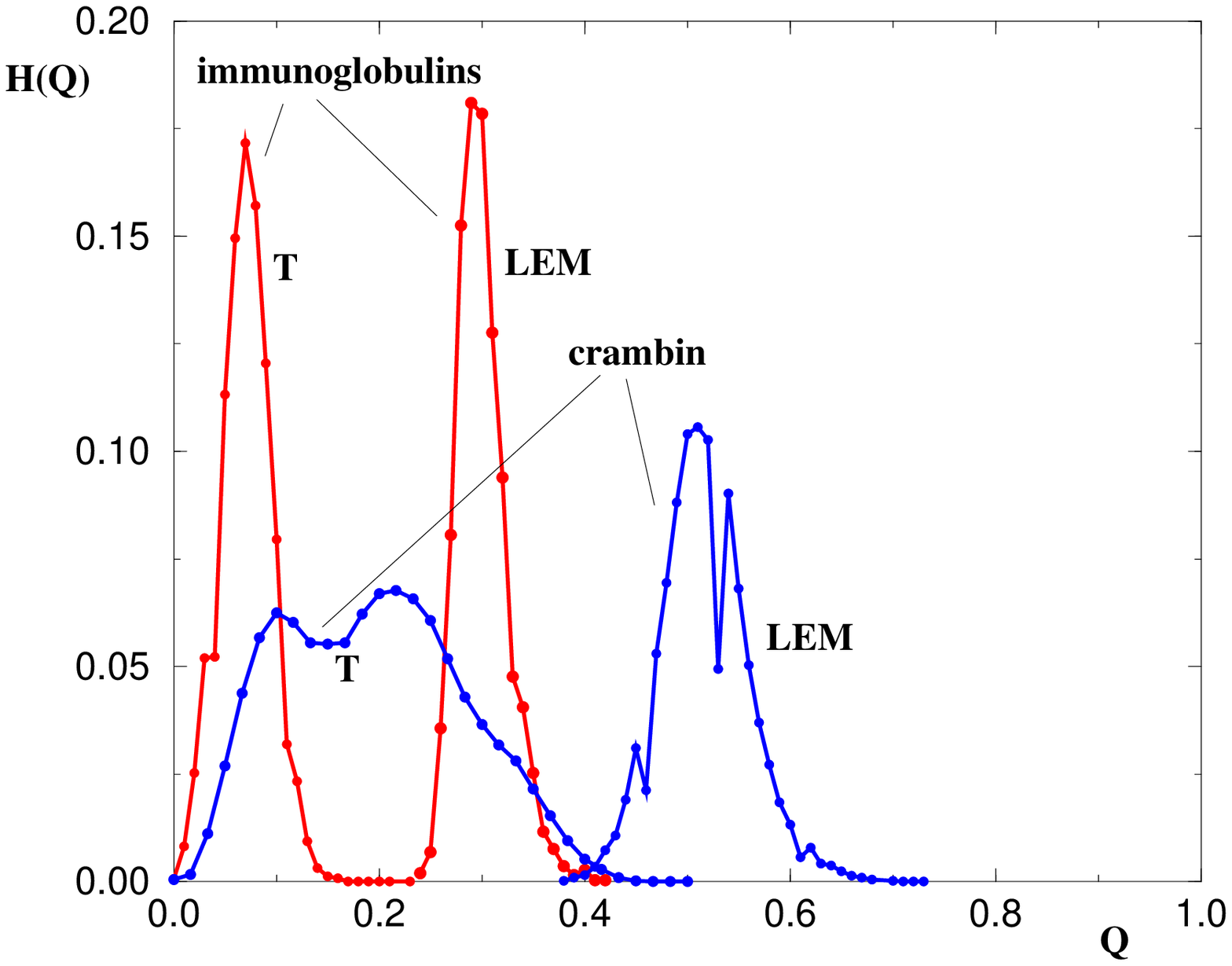,height=7.0cm,angle=0}
            \psfig{figure=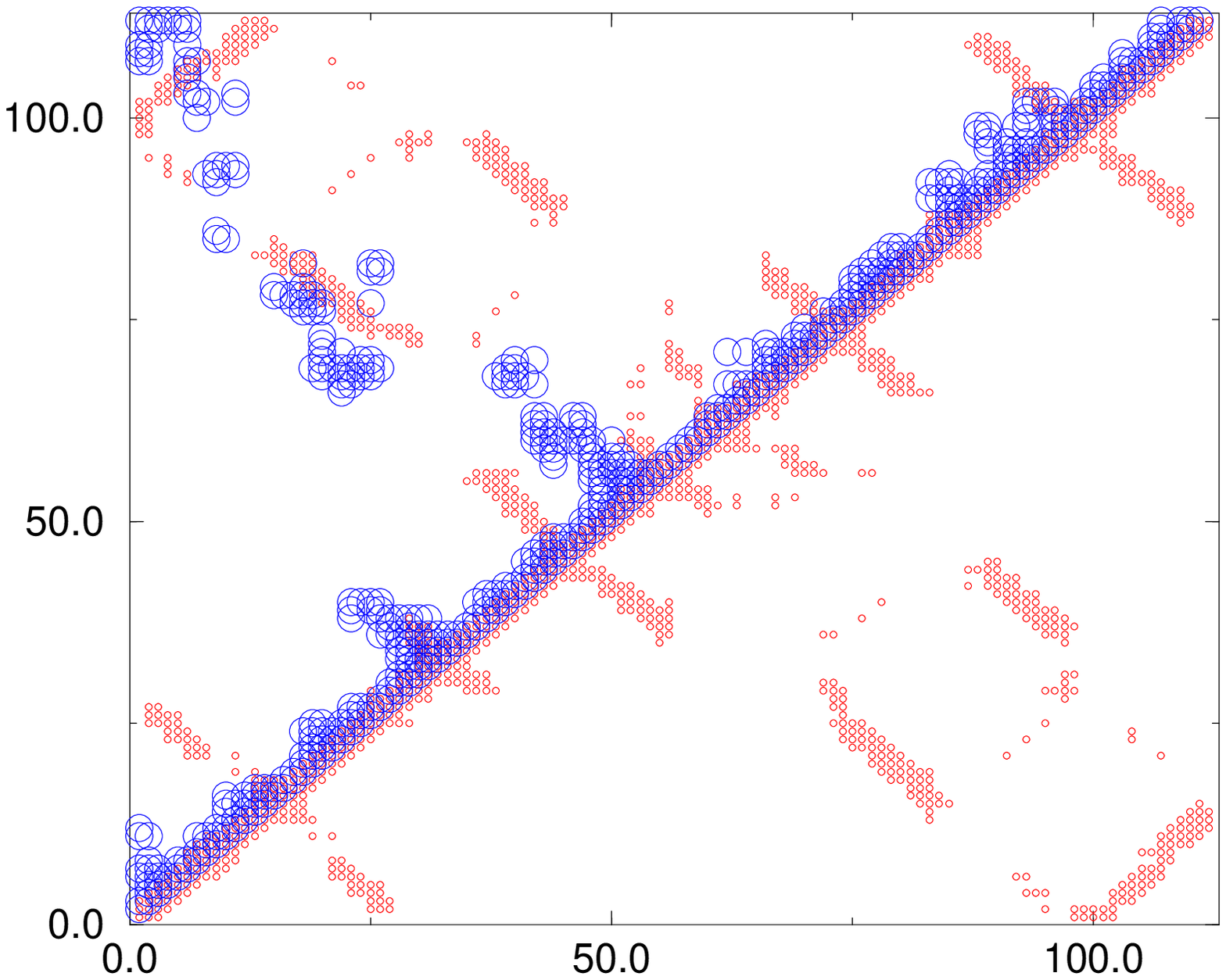,height=7.0cm,angle=0}}
\caption{
(a)
Histogram of the distribution of the overlap $Q$ between maps of low energy
(LEM) and the native map found in the
simulation with the hydrophobic term for immunoglobulins and for crambin.
For comparison, the distributions obtained by threading are also shown (T).
(b)
Typical low energy contact map (open circles) generated during the simulation
of 1dba. For comparison the native map is also shown (full dots).
}
\label{fig:ig}
\end{figure}

The failure to assign the lowest energy to the native contact map
is more severe in the case of immunoglobulins.
The low energy contact maps in the case of crambin did resemble
the native map to some extent, as shown in Fig. \ref{fig:qe}.
In the case of immunoglobulins there is very little
similarity between low energy maps and the native one (see Fig. \ref{fig:ig}b).

We think that this result deserves an explanation, which we propose in
the following argument. 
As already done for crambin, we obtained a set of decoys
by threading for immunoglobulins.
Such a distribution is narrower and shifted 
to the left with respect to the corresponding distribution for crambin. 
We assume that threading offers random protein-like conformations.
Therefore, the shift 
is originated by the exponential increase of the number 
of physical contact maps (Vendruscolo {\em et al.}, 1999) 
with the length of the proteins.
The improvement, as measured by the right-shift  of the distribution of $Q$,
obtained by learning energy parameters, is comparable
for crambin and for immunoglobulins. However, since immunoglobulins
are longer, one would have to work harder to obtain energy parameters
suitable for folding -- many more decoys are to be weeded out.
Since we proved that it is impossible to tune better the energy
parameters for both approximations of the energy that we discussed,
the only choice is to develop better forms for the energy.
This will be the object of future study.

\section{Final considerations}

We presented the results of our attempt to perform protein folding in
the space of contact maps. 

\begin{itemize}
\item
We demonstrated the feasibility of the original idea 
of performing energy minimization in contact map space.
\item
We proved that two simple phenomenological approximations to the free
energy can not possibly be tuned to assign the lowest energy
to the observed native conformation of even one single protein.
\item
We leave open the possibility to use better approximations for the
free energy and more detailed representations
of the chain underlying a contact map.
\item
We also leave open the possibility of predicting native states 
by other methodologies that do not use energy minimization.

\section*{Acknowledgments}
We are grateful to Ron Elber for discussing with us a similar approach,
based on Eq. \ref{eq:lowest} (unpublished), and to Gaddy Getz,
Ido Kanter, Edo Kussell, Rafael Najmanovich and Kibeom Park
for their contributions to different parts of the work presented.
This research was supported by grants from the Minerva Foundation,
the Germany-Israel Science Foundation (GIF) and by a grant from the
Israeli Ministry of Science.

\end{itemize}

\end{document}